\documentclass[twocolumn]{aastex62}
\pdfoutput=1 %for arXiv submission
\usepackage{amsmath,amstext}
\usepackage[T1]{fontenc}
\usepackage{apjfonts} 
\usepackage{natbib}
\usepackage{tabularx}
\usepackage{footnote}
\usepackage[flushleft, para]{threeparttable}

\shorttitle{[$\alpha$/Fe] from LRS in M31's Halo}
\shortauthors{Escala et al.}

\begin{document}
\title{Elemental Abundances in M31: Alpha and Iron Element Abundances from Low-Resolution Resolved Stellar Spectroscopy in the Stellar Halo}

\author{Ivanna Escala}
\thanks{E-mail: ie@astro.caltech.edu}
\affiliation{Department of Astronomy, California Institute of Technology, 1200 E California Blvd, Pasadena, CA, 91125, USA}
%\affiliation{National Science Foundation Graduate Research Fellow}
%\affiliation{Ford Foundation Predoctoral Fellow}
\author{Evan N. Kirby}
\affiliation{Department of Astronomy, California Institute of Technology, 1200 E California Blvd, Pasadena, CA, 91125, USA}
\author{Karoline M. Gilbert}
\affiliation{Space Telescope Science Institute, 3700 San Martin Drive, Baltimore, MD 21218, USA}
\affiliation{Department of Physics \& Astronomy, Bloomberg Center for Physics and Astronomy, John Hopkins University, 3400 N. Charles St, Baltimore, MD 21218, USA}
\author{Emily C. Cunningham}
\affiliation{Department of Astronomy and Astrophysics, University of California, Santa Cruz, 1156 High St, Santa Cruz, CA, 95064, USA}
%\author{Gina Duggan}
%\affiliation{Department of Astronomy, California Institute of Technology, 1200 E California Blvd, Pasadena, CA, 91125, USA}
\author{Jennifer Wojno}
\affiliation{Department of Physics \& Astronomy, Bloomberg Center for Physics and Astronomy, John Hopkins University, 3400 N. Charles St, Baltimore, MD 21218, USA}

\received{2018 November 15}
\revised{2019 April 9}
\accepted{2019 April 30}

\begin{abstract}
%Stellar halos provide a record of the earliest stages of a galaxy's formation as well as its mass assembly. 
Measurements of [Fe/H] and [$\alpha$/Fe] can probe the minor merging history of a galaxy, providing a direct way to test the hierarchical assembly paradigm. 
%The stellar halo and tidal streams of M31 provide a complement to the same structures around the Milky Way (MW).
While measurements of [$\alpha$/Fe] have been made in the stellar halo of the Milky Way, little is known about detailed chemical abundances in the stellar halo of M31. To make progress with existing telescopes, we apply spectral synthesis to low-resolution DEIMOS spectroscopy (R $\sim$ 2500 at 7000 \AA) across a wide spectral range (4500~\AA \ $<$ $\lambda$ $<$ 9100~\AA). By applying our technique to low-resolution spectra of \deleted{135}\added{170} \deleted{red giant branch (RGB)} \added{giant} stars in \deleted{4}\added{5} MW globular clusters, we demonstrate that our technique reproduces previous measurements from higher resolution spectroscopy. Based on the intrinsic dispersion in [Fe/H] and [$\alpha$/Fe] of individual stars in our combined cluster sample, we estimate systematic uncertainties of $\sim$0.11 dex and $\sim$\deleted{0.04}{0.09} dex in [Fe/H] and [$\alpha$/Fe], respectively.
%over a more limited spectral range (6300 - 9100 \AA) using the DEIMOS 1200G grating.
We apply our method to deep, low-resolution spectra of \deleted{14}\added{11} \deleted{RGB}\added{red giant branch} stars in the smooth halo of M31, resulting in higher signal-to-noise per spectral resolution element compared to DEIMOS medium-resolution spectroscopy, given the same exposure time and conditions. We find $\langle$[$\alpha$/Fe]$\rangle$ = 0.49 $\pm$ \deleted{0.31}\added{0.29} dex and $\langle$[Fe/H]$\rangle$ = $-$\deleted{1.53}\added{1.59} $\pm$ \deleted{0.52}\added{0.56} dex for our sample. This implies that\added{---much like the Milky Way---}the smooth halo field \added{of M31} is likely composed of disrupted dwarf galaxies with truncated star formation histories that were accreted early in \deleted{the formation history of M31}\added{the halo's formation.}
%For the first time, we present measurements of [Fe/H] and [$\alpha$/Fe] of sufficient quality 
%and sample size 
%to construct quantitative models of galactic chemical evolution in the M31 system. 
\end{abstract}

\keywords{stars: abundances -- galaxies: abundances -- galaxies: halos -- galaxies: formation  -- Local Group}

\section{Introduction}

Stellar chemical abundances are a key component in determining the origins of stellar halos of Milky Way (MW) like galaxies, providing insight into the formation of galaxy-scale structure. The long dynamical times of stellar halos allow tidal features to remain identifiable in phase space, in terms of kinematics and chemical abundances, for Gyr timescales. Stellar chemical abundances of stars retain information about star formation history and accretion times of progenitor satellite galaxies, even when substructures can no longer be detected by kinematics alone. In particular, measurements of metallicity\footnote{We define metallicity in terms of stellar iron abundance, [Fe/H], where [Fe/H] = $\log_{10}(n_\textrm{Fe}/n_\textrm{H}) - \log_{10}(n_\textrm{Fe}/n_\textrm{H})_\odot$} and $\alpha$-element abundances provide a way of directly testing the hierarchical assembly paradigm central to $\Lambda$CDM cosmology, providing a fossil record of the formation environment of stars accreted onto the halo.

The [$\alpha$/Fe] ratio serves as a useful diagnostic of formation history, given that it traces the star formation timescales of a galaxy (e.g., \citealt{GilmoreWyse1998}). Type II supernovae (SNe) produce abundant $\alpha$-elements (O, Ne, Mg, Si, S, Ar, Ca, and Ti), 
increasing [$\alpha$/Fe], whereas Type Ia SNe produce Fe-rich ejecta, reducing [$\alpha$/Fe]. While measurements of [$\alpha$/Fe] have been made in the stellar halo of the MW, little is known about the detailed chemical abundances of the stellar halo of M31.  A comparable understanding of the properties of the MW and M31 stellar halos is required to verify basic assumptions about how the MW evolved, where such assumptions are used to extrapolate MW-based results to studies of galaxies beyond the Local Group.

Although high-resolution ($R$ $\gtrsim$ 15,000), high signal-to-noise (S/N) spectra enables simultaneous measurements of a star's temperature, surface gravity, and individual element abundances based on individual lines, it is impractical to achieve high enough S/N for traditional high-resolution spectroscopic abundance analysis (e.g., \citealt{KirbyCohen2012}) for red giant branch (RGB) stars at the distance of M31 (783 kpc; \citealt{StanekGarnavich1998}). 

It is possible to obtain spectroscopic metallicity measurements of M31 RGB stars from medium-resolution spectra ($R$ $\sim$ 6000) using spectral synthesis (e.g., \citealt{Kirby2008}). This method leverages the entire spectrum's metallicity information simultaneously, enabling measurements of abundances from relatively low S/N spectra. \citet{Kirby2008b,Kirby2010MULTI-ELEMENT,Kirby2013} successfully measured [Fe/H] and [$\alpha$/Fe] in MW globular clusters (GCs), MW dwarf spheroidal (dSph) satellite galaxies, and Local Group dwarf irregular galaxies, showing that abundances can be measured to a precision of $\sim$ 0.2 dex from spectra with S/N $\sim$ 15~\AA$^{-1}$. 

Only in 2014 has spectral synthesis been applied to individual RGB stars in the M31 system for the first time \citep{Vargas2014a,Vargas2014b}. Existing spectroscopic chemical abundance measurements in M31 are primarily based on metallicity estimates from the strength of the calcium triplet \citep{Chapman2006,Koch2008,Kalirai2009,Richardson2009,Tanaka2010,Ibata2014,Gilbert2014,Ho2015}. \citet{Vargas2014a} measured [$\alpha$/Fe] and [Fe/H] for a total of 226 red giants in 9 M31 satellite galaxies. Although \citet{Vargas2014a} measured [$\alpha$/Fe] for 9 M31 dSphs, only a single dSph, And V, shows a clear chemical abundance pattern, where [$\alpha$/Fe] declines with [Fe/H]. However, the present spectroscopic sample size and measurement uncertainties of the And V data enable only qualitative conclusions about the chemical evolution of the dSph. Obtaining more quantitative descriptions of the chemical enrichment and star formation histories of the M31 system requires higher S/N spectroscopic data, which results in smaller uncertainties on abundance measurements. Only then can one-zone numerical chemical evolution models \citep{Lanfranchi2003,Lanfranchi2007,Lanfranchi2010,Lanfranchi2006,Kirby2011a} be reliably applied to measurements to derive star formation and mass assembly histories. 

%Using the DEIMOS (DEep Imaging Multi-Object Spectrograph; \citealt{Faber2003}) 1200 line mm$^{-1}$ ($R$ $\sim$ 7000 at the Ca triplet, $\lambda$ $\sim$ 8500~\AA) grating for medium-resolution spectrsocopy (MRS), the SPLASH (Spectroscopic and Photometric Landscape of Andromeda's Stellar Halo) \citep{Kalirai2010,Gilbert2012,Ho2012, Tollerud2012,Gilbert2014,Gilbert2018} collaboration obtained spectra for 50 fields in the M31 stellar halo, observing RGB stars for $\sim$ 1 hour in the wavelength range $\lambda$ $\sim$ 6300 - 9100~\AA. The median S/N is $\sim$5.7~\AA$^{-1}$ \citep{Gilbert2014}, which is insufficient to measure full abundance distributions.
%Only a small fraction of stars had S/N $\gtrsim$ 15~\AA$^{-1}$ sufficient to measure full abundance distributions. 
Although \citet{Vargas2014a,Vargas2014b} demonstrated the feasibility of measuring [Fe/H] and [$\alpha$/Fe] at the distance of M31, measuring [Fe/H] and [$\alpha$/Fe] more precisely requires deep ($\sim$6 hour) observations with DEIMOS using the 600 line mm$^{-1}$ grating to yield higher S/N for the same exposure time and observing conditions. For magnitudes fainter than $I_0$ $\sim$ 21 (0.5 magnitudes below the tip of M31's RGB), sky line subtraction at $\lambda$ $>$ 7000 \AA \ becomes the dominant source of noise in DEIMOS spectra observed with the 1200 line mm$^{-1}$ grating. Given the access to blue optical wavelengths granted by the 600 line mm$^{-1}$ grating, its spectra are less susceptible to the effects of sky noise. Additionally, using the 600 line mm$^{-1}$ grating achieves higher S/N per pixel for stars as faint as $I_0$ $\sim$ 21.8.

%For background-limited spectra with low S/N (S/N $\lesssim$ 8 \AA$^{-1}$), spectral coaddition similar stars is a viable method to measure abundances. This is particularly true for [Fe/H], where color and luminosity can be used as a proxy for the effective temperature and surface gravity of stars in the binning stage of the coaddition process (e.g., \citealt{Yang2013}). In the case of $\alpha$-element abundances, photometric constraints are not available, such that coaddition involves an inherently heterogeneous group of stars. Although \citet{Yang2013} concluded that coaddition of medium-resolution spectra can reliably be applied to M31 RGB stars to obtain detailed chemical abundance patterns, their measurements of $\alpha$-elements from spectral coaddition show large scatter compared to expectations. 

%If feasible, measuring abundances from individual stellar spectra is thus preferred. 
Although using the 600 line mm$^{-1}$ grating with DEIMOS results in a gain in S/N and wavelength coverage, it corresponds to a decrease in spectral resolution ($\sim$2.8 \AA \ FWHM, or R $\sim$ 2500 at 7000 \AA, compared to $\sim$1.3 \AA \ and R $\sim$ 5400 for 1200 line mm$^{-1}$). Increasing the spectral range compensates for the decrease in spectral resolution, given the increase in the amount of available abundance information contained in the spectrum resulting from the higher density of absorption features at bluer optical wavelengths.

In this paper, we expand upon the technique first presented by \citet{Kirby2008}, applying spectral synthesis to low-resolution spectroscopy (LRS; R $\sim$ 2500) across a wide spectral range ($\lambda$ $\sim$ 4500 - 9100 \AA). In \S~\ref{sec:obs}, we describe our data reduction and GC observations. \S~\ref{sec:prep_spec} and \S~\ref{sec:abund} detail our preparations to the observed spectrum and the subsequent abundance analysis. This includes a presentation of our new line list and grid of synthetic spectra. In \S~\ref{sec:results}, we illustrate the efficacy of our technique applied to MW GCs and compare our results to chemical abundances from high-resolution spectroscopy in \S~\ref{sec:hrs}. We quantify the associated systematic uncertainties in \S~\ref{sec:sys}. We conclude by measuring [$\alpha$/Fe] and [Fe/H] in a M31 stellar halo field in \S~\ref{sec:m31} and summarize in \S~\ref{sec:summary}.
%For the first time, we present measurements of [$\alpha$/Fe], while simultaneously measuring [Fe/H] to high precision, of sufficient quality to construct quantitative models of chemical evolution in the M31 satellite system. 

\section{Observations}
\label{sec:obs}

\begin{table*}
\centering
%\begin{threeparttable}
\caption{MW and dSph DEIMOS Observations}
\begin{tabular}{lccccccccc}
\hline
\hline
Object & Slitmask & $\alpha_\textrm{J2000}$ & $\delta_\textrm{J2000}$ & Date & Seeing ('') & Airmass & $t_{\textrm{exp}}$\footnote{In the case of multiple exposures with unequal exposure time, we indicate the total exposure time.}  (s) & $N_\textrm{target}$ & $N_\textrm{member}$\footnote{Number of RGB and AGB members \citep{Kirby2016} per slitmask. For dSph slitmasks, we do not fully evaluate membership since we utilized these observations only for comparison to the HRS literature (\S~\ref{sec:hrs}).}\\
\hline
\multicolumn{10}{c}{MW Globular Clusters} \\ \hline
NGC 2419\dotfill & n2419c & 07:38:09.67 & +38:51:15.0 & 2015 Oct 9 & 0.6 & 1.23 & 2 $\times$ 1380 & 92 & 61\\
NGC 1904 (M79) & 1904l2 & 05:24:15.37 & $-$24:31:31.3 & 2015 Oct 8 & 0.8 & 1.40 & 2 $\times$ 1260 & 96 & 18\\
NGC 6864 (M75) & n6864a & 20:06:14.03 & $-$21:55:16.4 & 2015 May 19 & 0.9 & 1.56 & 3 $\times$ 1080 & 86 & 35\\
%NGC 6656 (M22) & 2015 May 18 & 0.9 & 1.44 & 3 $\times$ 900 & 64\\
NGC 6341 (M92) & n6341b & 17:17:23.68 & +43:06:49.4 & 2018 Oct 11 & 0.6 & 1.52 & 6 $\times$ 300 & 146 & 33\\
NGC 7078 (M15) & 7078l1 & 21:29:48.03 & +12:10:23.0 & 2015 May 19 & 0.9 & 1.16 & 2 $\times$ 1140 & 169 & 48 \\ \hline
\multicolumn{10}{c}{dSphs} \\ \hline
Draco\dotfill & dra11 & 17:19:46.87 & +57:57:21.1 & 2019 Mar 10 & 1.6 & 1.39 & 4280 & 138 & \nodata \\ 
Canes Venatici I & CVnIa & 13:28:02.47 & +33:32:49.5 & 2018 May 20 & 1.0 & 1.3 & 6 $\times$ 1200 & 122 & \nodata \\
Ursa Minor\dotfill & bumib & 15:09:28.75 & +67:13:07.1 & 2018 May 20 & 1.0 & 1.68 & 7 $\times$ 1200 & 124 & \nodata\\
\hline
\multicolumn{10}{c}{MW Halo} \\ \hline
HD20512\dotfill & LVMslitC & 03:18:27.14 & +15:10:38.29 & 2019 Mar 10 & 1.2 & 1.35 & 10 & \nodata & \nodata\\
HD21581\dotfill & LVMslitC & 03:28:54.48 & $-$00:25:03.10 & 2019 Mar 10 & 1.2 & 1.45 & 171 & \nodata & \nodata\\
HD88609\dotfill & LVMslitC & 10:14:28.98 & +53:33:39.34 & 2019 Mar 10 & 1.2 & 1.43 & 44 & \nodata & \nodata\\
SAO134948\dotfill & LVMslitC & 06:46:03.69 & +67:13:45.75 & 2019 Mar 10 & 1.2 & 1.49 & 180 & \nodata & \nodata\\
BD+80245\dotfill & LVMslitC & 08:11:06.23 & +79:54:29.55 & 2019 Mar 10 & 1.2 & 1.99 & 3 $\times$ 300 & \nodata & \nodata\\
\hline
\end{tabular}
\label{tab:obs}
%\begin{tablenotes}
%\item[a] Seeing in arcseconds, \item[b] Airmass, \item[c] Total exposure time, \item[d] Number of RGB and AGB members per slitmask.
%\end{tablenotes}
%\end{threeparttable}
\end{table*}

%We have measured [Fe/H] and [$\alpha$/Fe] using the DEIMOS 600 line mm$^{-1}$ grating. For S/N $\sim$ 30 \AA$^{-1}$, we can obtain metallicities at precisions as high as 0.13 dex. 
We utilize observations of Galactic GCs, \added{MW dwarf spheroidal (dSph) galaxies, and MW halo stars} (Table~\ref{tab:obs}) taken using Keck/DEIMOS \citep{Faber2003} to validate our LRS method of spectral synthesis. For our science configuration (\deleted{both for MW GCs and}\added{for all observations, including} M31 observations, \S~\ref{sec:m31_obs}), we used the GG455 filter with a central wavelength of 7200 \AA, in combination with the 600ZD grating and 0.7'' slitwidths. \added{When targeting individual stars, such as the MW halo stars in Table~\ref{tab:obs}, we utilized the longslit, as opposed to a slitmask intended to target multiple stars simultaneously.}

The spectral resolution \added{for the 600 line mm$^{-1}$ grating} is approximately $\sim$2.8 \AA \ FWHM, compared to $\sim$1.3 \AA \ FWHM for the 1200 line mm$^{-1}$ grating used in prior observations \citep{Kirby2010MULTI-ELEMENT,Kirby2013}. The wavelength range for each spectrum obtained with the 600ZD grating is within 4100 \AA \ $-$ 1 $\mu$m, where we generally omit $\lambda$ $\lesssim$ 4500 \AA, owing to poor S/N in this regime and the presence of the G band. We also omit $\lambda$ > 9100 \AA, which extends beyond the wavelength coverage of our grid of synthetic spectra (\S~\ref{sec:synth_spec}).

To extract one-dimensional spectra from the raw DEIMOS data, we used a modification of version 1.1.4 of the data reduction pipeline developed by the DEEP2 Galaxy Redshift Survey \citep{Cooper2012,Newman2013}. \citet{Guhathakurta2006} provides a detailed description of the data reduction process. Modifications to the software include those of \citet{SimonGeha2007}, where the pipeline was re-purposed for bright unresolved stellar sources (as opposed to faint, resolved galaxies). In addition, we include custom modifications to correct for atmospheric refraction in the two-dimensional raw spectra, which affects bluer optical wavelengths, and to identify lines in separate arc lamp spectra, as opposed to a single stacked arc lamp spectrum.

\section{Preparing the Spectrum for Abundance Measurement}
\label{sec:prep_spec}

\subsection{Telluric Absorption Correction}
\label{sec:telluric}

Unlike the red side of the optical (6300 - 9100 \AA), there is no strong telluric absorption in the bluer regions (4500 - 6300 \AA). As such, we do not make any corrections to the observed stellar spectra to take into account absorption from Earth's atmosphere in this wavelength range.

For the red (6100 - 9100 \AA), we correct for the absorption of Earth's atmosphere using the procedure described in \citet{Kirby2008}. We adopt HD066665 (B1V), observed on April 23, 2012 with an airmass of 1.081, using a long slit in the same science configuration (\S~\ref{sec:obs}) as our data, as our spectrophotometric standard.

\subsection{Spectral Resolution Determination}

In contrast to \citet{Kirby2008}, who determined the spectral resolution as a function of wavelength based on the Gaussian widths of hundreds of sky lines, we assume a constant resolution, expressed as the typical FWHM of an absorption line, across the entire observed spectrum ($\sim$4500$-$9100 \AA). Owing to the dearth of sky lines at bluer wavelengths, the number of available sky lines is insufficient to reliably determine the resolution as a function of wavelength. As an alternative, we introduce an additional parameter, $\Delta\lambda$, or the spectral resolution, into our chi-squared minimization, which determines the best-fit synthetic spectrum for each observed spectrum (\S~\ref{sec:spec_abund}).
%We then fix the spectral resolution of each star in a given mask to the average value determined from the entire mask (~\S~\ref{sec:spec_abund}).

\subsection{Continuum Normalization}
\label{sec:cont_norm}

It is necessary to normalize the observed flux by its slowly varying stellar continuum in order to meaningfully compare to synthetic spectra for the abundance determination (\S~\ref{sec:spec_abund}). To obtain reliable abundances from spectral synthesis of low- and medium-resolution spectra dominated by weak absorption features, the continuum determination must be accurate \citep{Shetrone2009,Kirby2009}. This is particularly important for bluer wavelengths, where absorption lines are so numerous and dense that we cannot define ``continuum regions'' \citep{Kirby2008} in the blue. Instead, we utilize the entire observed spectrum, excluding regions with strong telluric absorption and bad pixels, to determine the continuum for 4500 $-$ 9100 \AA. In contrast to \citet{Kirby2008}, we do not utilize continuum regions at redder wavelengths (6300 - 9100 \AA), despite the fact that they can be reliably defined, to maintain consistency in the continuum normalization method between each wavelength region of the observed spectrum.

We determined the initial continuum fit to the raw observed spectrum, which we shift into the rest frame, using a third-order B-spline with a breakpoint spacing of 200 pixels, excluding 5 pixels around the chip gap and at the start and stop wavelengths of the observed spectrum. In all steps, we weighted the spline fit by the inverse variance of each pixel in the observed spectrum. We performed sigma clipping, such that pixels that deviate by more than 5$\sigma$ (0.1$\sigma$) above (below) the fit are excluded from the subsequent continuum determination, where $\sigma$ is the inverse square root of the inverse variance array. 
%where $\sigma$ is the standard deviation of the differences between the pixels and the B-spline fit.
We did not perform the fit iteratively beyond the above steps, given that our stringent criterion to prevent the numerous absorption lines from offsetting our continuum determination can eliminate a significant fraction of the pixels from subsequent iterations of the fit.

%We employ a similar procedure for redder wavelengths (6300 - 9100 \AA), except we implement symmetric sigma clipping (5$\sigma$ above and below the fit) and use only continuum regions in the fit \citep{Kirby2008}, which we re-define at the spectral resolution of the 600ZD grating (Gaussian width 1.2 \AA).

To further refine our continuum determination, we recalculate the continuum fit iteratively in the initial step of the abundance analysis (\S~\ref{sec:spec_abund}). Once we have found a best-fit synthetic spectrum, we divide the continuum-normalized observed spectrum by the best-fit synthetic spectrum to construct a ``flat noise'' spectrum, which captures the higher order terms in the observed spectrum not represented in the fit. We fit a third-order B-spline with a breakpoint spacing of 100 pixels to the flat noise spectrum, excluding 3$\sigma$ deviant (above and below the fit) pixels, dividing the continuum-normalized observed spectrum by this fit. The modified continuum-normalized spectrum is then used in the next iteration of the continuum refinement until convergence is achieved (\S~\ref{sec:spec_abund}).

\subsection{Pixel Masks}
\label{sec:custom_mask}

\begin{table}
\centering
\caption{Spectral Features (4100 - 6300 \AA)}
\label{tab:spec_features}
\begin{tabular}{lc}
\hline
\hline
Feature & Wavelength(s) (\AA)\\
\hline
H$\delta$ & 4101.734\\
Ca I & 4226.730\\
G band (CH absorption) & 4300-4315 \\
H$\gamma$ & 4340.462 (4335-4345)\footnote{Wavelength regions indicated in paranthesis indicate regions that are omitted from the spectral fit.}\\
H$\beta$ & 4861.35 (4856-4866)\\
Mg I (b4) & 5167.322\\
Mg I (b2) & 5172.684\\
Mg I (b1) & 5183.604\\
Mg H & 4845,5622\\
Na D1,D2 & 5895.924,5889.951 (5885-5905)\\
\hline
\end{tabular}
\end{table}

In addition to wavelength masks corresponding to a particular abundance (\S~\ref{sec:spec_regions}), we constructed a pixel mask for each analyzed observed spectrum. Typically excluded regions include 5 pixels on either side of the chip gap between the blue and red sides of the CCD, areas with improper sky line subtraction, the region around the Na D1 and D2 lines (5585 - 5905 \AA), and other apparent instrumental artifacts. Table~\ref{tab:spec_features} includes a summary of prominent spectral features in DEIMOS spectra between 4100 - 6300 \AA, where wavelength ranges given in parenthesis indicate regions that are masked. For example, we excluded 10 \AA \ regions around H$\gamma$ (4335 - 4345 \AA) and H$\beta$ (4856 - 4866 \AA). \added{Given that MOOG \citep{1973PhDT.......180S}, the spectral synthesis software utilized to generate our grid of synthetic spectra (\S~\ref{sec:synth_spec}}, does not incorporate the effects of non-local thermodynamic equilibrium, \added{it} cannot properly model the strong Balmer lines. If necessary, we also masked regions where the initial continuum fit failed, most often owing to degrading signal-to-noise as a function of wavelength at bluer wavelengths ($\lesssim$ 4500 \AA). As for the red (6300 - 9100 \AA), we adopted the pixel mask from \citet{Kirby2008}, which excludes spectral features such as the Ca II triplet, H$\alpha$, and regions with strong telluric absorption.

\subsection{Signal-to-Noise Estimation}
\label{sec:sn}

We estimate S/N per Angstrom for objects observed with the 600 line mm$^{-1}$ grating from wavelength regions of the spectrum utilized in the initial continuum determination (\S~\ref{sec:cont_norm}). Given that we cannot define continuum regions for wavelengths blueward of 6300 \AA, we calculate the S/N after the continuum refinement process (\S~\ref{sec:spec_abund}). We estimate the noise as the deviation between the continuum-refined observed spectrum and the best-fit synthetic spectrum and the signal as the best-fit synthetic spectrum itself. The S/N estimate per pixel is the \deleted{mean}\added{median} of the S/N as a function of wavelength calculated from these quantities, where we exclude pixels that exceed the average noise threshold by more than 3$\sigma$. To convert to units of per Angstrom, we multiply this quantity by the inverse square root of the pixel scale ($\sim$ 0.64 \AA \ for the 600 line mm$^{-1}$ grating). 

\begin{table}
\begin{threeparttable}
%\centering
\caption{Blue Line List (4100 - 6300 \AA)}
\label{tab:linelist}
\begin{tabular*}{\columnwidth}{l @{\extracolsep{\fill}} ccc}
\hline
\hline
Wavelength (\AA) & Species\tnote{a} & EP\tnote{b} (eV) & $\log$ $gf$\tnote{c}\\
\hline
5183.409 & 57.1 & 0.403 & -0.6\\
5183.414 & 69.1 & 4.744 & -2.65\\
5183.436 & 24.1 & 6.282 & -3.172\\
5183.465 & 26.0 & 3.111 & -5.06\\
5183.466 & 27.0 & 4.113 & -1.187\\
5183.493 & 106.00112 & 3.244 & -2.848\\
5183.506 & 106.00113 & 1.569 & -3.974\\
5183.518 & 607.0 & 1.085 & -4.211\\
5183.544 & 26.0 & 5.064 & -3.886\\
5183.55 & 58.1 & 1.706 & -2.27\\
5183.565 & 106.00112 & 3.55 & -2.811\\
5183.578 & 607.0 & 1.204 & -2.653\\
5183.598 & 23.1 & 6.901 & -3.568\\
5183.604 & 12.0 & 2.717 & -0.167\\
5183.615 & 607.0 & 1.085 & -3.003\\
5183.683 & 607.0 & 1.205 & -4.499\\
5183.683 & 106.00113 & 1.29 & -4.921\\
5183.686 & 106.00113 & 2.371 & -3.586\\
5183.708 & 40.0 & 0.633 & -1.62\\
5183.709 & 22.1 & 1.892 & -2.535\\
5183.748 & 58.1 & 1.482 & -1.56\\
5183.794 & 106.00112 & 1.43 & -2.866\\
5183.803 & 24.0 & 5.277 & -3.52\\
\hline
\end{tabular*}
\begin{tablenotes}
\item Note. \textemdash The line list presented here is a subset of the entire line list, which spans 4100 - 6300 \AA. The range of wavelengths presented here spans 0.4 \AA \ around the strong Mg I line at 5183.604 \AA. The line list is formatted to be compatible with MOOG. 
\item[a] A unique code corresponding to chemical species. For example, 12.0 indicates Mg I, 22.1 indicates Ti II, 106.00112 indicates CN for carbon-12, and 106.0113 indicates CN for carbon-13.
\item[b] Excitation potential.
\item[c] Oscillator strength. Transitions modified in the vetting process have less than three decimal places.
\end{tablenotes}
\end{threeparttable}
\end{table}

\section{Chemical Abundance Analysis}
\label{sec:abund}

\begin{figure*}
\centering
\includegraphics[width=\textwidth]{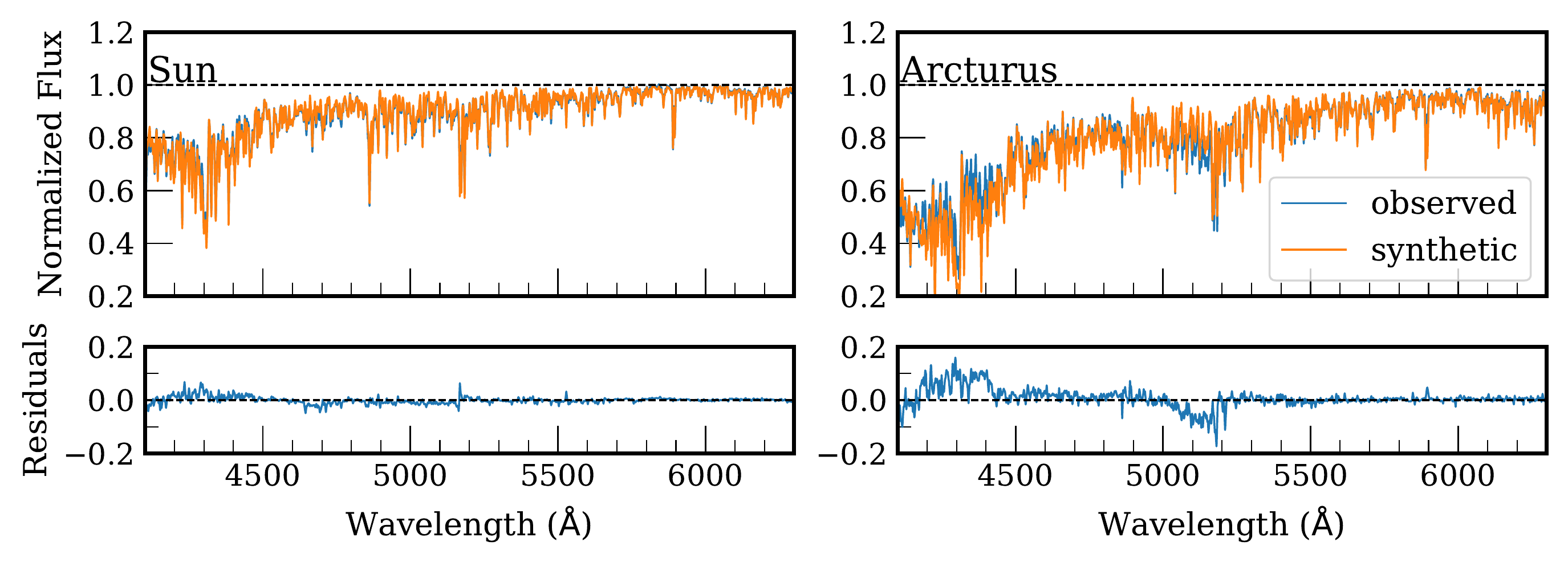}
\caption{(\textit{Top}) A comparison between high-resolution spectra \citep{2000vnia.book.....H} (\textit{blue}) of the Sun (\textit{left}) and Arcturus (\textit{right}) and synthetic spectra (\textit{orange}) generated using the blue line list (\S~\ref{sec:line_list}). Both spectra are smoothed to the expected resolution of the DEIMOS 600ZD grating ($\sim$ 2.8 \AA). For a description of synthetic spectrum generation, see \S~\ref{sec:synth_spec}. (\textit{Bottom}) The difference between the observed and synthetic spectra for the Sun and Arcturus. To improve the agreement between the synthetic and observed spectra, we have manually vetted the line list, adjusting the oscillator strengths of discrepant atomic transitions as necessary. In this process, we have favored the Sun over Arcturus, thus the larger residuals between the observed and synthetic spectra for the latter star (which has a lower effective temperature).
\label{fig:vetting}}
\end{figure*}

Here, we present a new library of synthetic spectra in the range 4100 - 6300~\AA. In this section, we describe our procedure for spectral synthesis in the blue, where we use our new grid in conjunction with the red grid of \citet{Kirby2008} to measure abundances across an expanded optical range (4100 - 9100 \AA).

\subsection{Line List}
\label{sec:line_list}

We constructed a line list of wavelengths, excitation potentials (EPs), and oscillator strengths ($\log$ $gf$) for atomic and molecular transitions in the spectral range covering 4100 - 6300~\AA \ for stars in our stellar parameter range ($T_\textrm{eff}$ $>$ 4000 K). We queried the Vienna Atomic Line Database (VALD; \citealt{Kupka1999}) and the National Institute of Standards and Technology (NIST) Atomic Spectra Database \citep{Kramida2016NIST5.4} for all transitions of neutral or singly ionized atoms with EP $<$ 10 eV and $\log$ $gf$ $>$ -5, supplementing the line list with molecular \citep{1992RMxAA..23...45K} and hyperfine transitions \citep{1993PhST...47..110K}. All \ion{Fe}{1} line oscillator strengths from \citet{FuhrWiese2006} are included in the NIST database. 

Next, we compared synthetic spectra (\S~\ref{sec:synth_spec}) of the Sun and Arcturus, generated from our line list and model stellar atmospheres, to high resolution spectra \citep{2000vnia.book.....H} of the respective stars. We adopted $T_\textrm{eff}$ = 5780 K, $\log$ $g$ = 4.44 dex, [Fe/H] = 0 dex, and [$\alpha$/Fe] = 0 dex  for the Sun. For Arcturus, we adopted $T_\textrm{eff}$ = 4300 K, $\log$ $g$ = 1.50 dex, and [Fe/H] = -0.50 dex, and [$\alpha$/Fe] = 0 dex \citep{1993ApJ...404..333P}. 

To produce agreement between the synthetic and observed spectra, we vetted the line list by manually adjusting the oscillator strengths of aberrant atomic lines as necessary. We preferred the Sun over Arcturus in this process, given that Arcturus is a cool \added{K-}giant \added{star} with stronger molecular absorption features (e.g., the G band) that are more difficult to match. For features absent from the line list, which could not be resolved by considering lines with $\log$ $gf$ $<$ $-$5, we included \ion{Fe}{1} transitions with EPs and $\log$ $gf$ to match the observed strength in both the Sun and Arcturus. \added{We present} the final blue line list \added{in a format compatible with MOOG in Table~\ref{tab:linelist}}. \added{The line list} contains 132 chemical species (atomic, molecular, neutral, and ionized), including 74 unique elements and 2 molecules (CN and CH). In total, the line list contains 53,164 atomic line transitions and 58,062 molecular transitions. 

Figure~\ref{fig:vetting} illustrates a comparison between the Hinkle spectra and their syntheses for the Sun and Arcturus. At the expected resolution of the DEIMOS 600ZD grating ($\sim$ 2.8 \AA), the mean absolute deviation of the residuals between the observed spectra and their syntheses across the wavelength range of the line list are 8.3 $\times$ 10$^{-3}$ and 2.2 $\times$ 10$^{-2}$ for the Sun and Arcturus respectively.

\subsection{Synthetic Spectra}
\label{sec:synth_spec}

\begin{figure*}
\centering
\includegraphics[width=\textwidth]{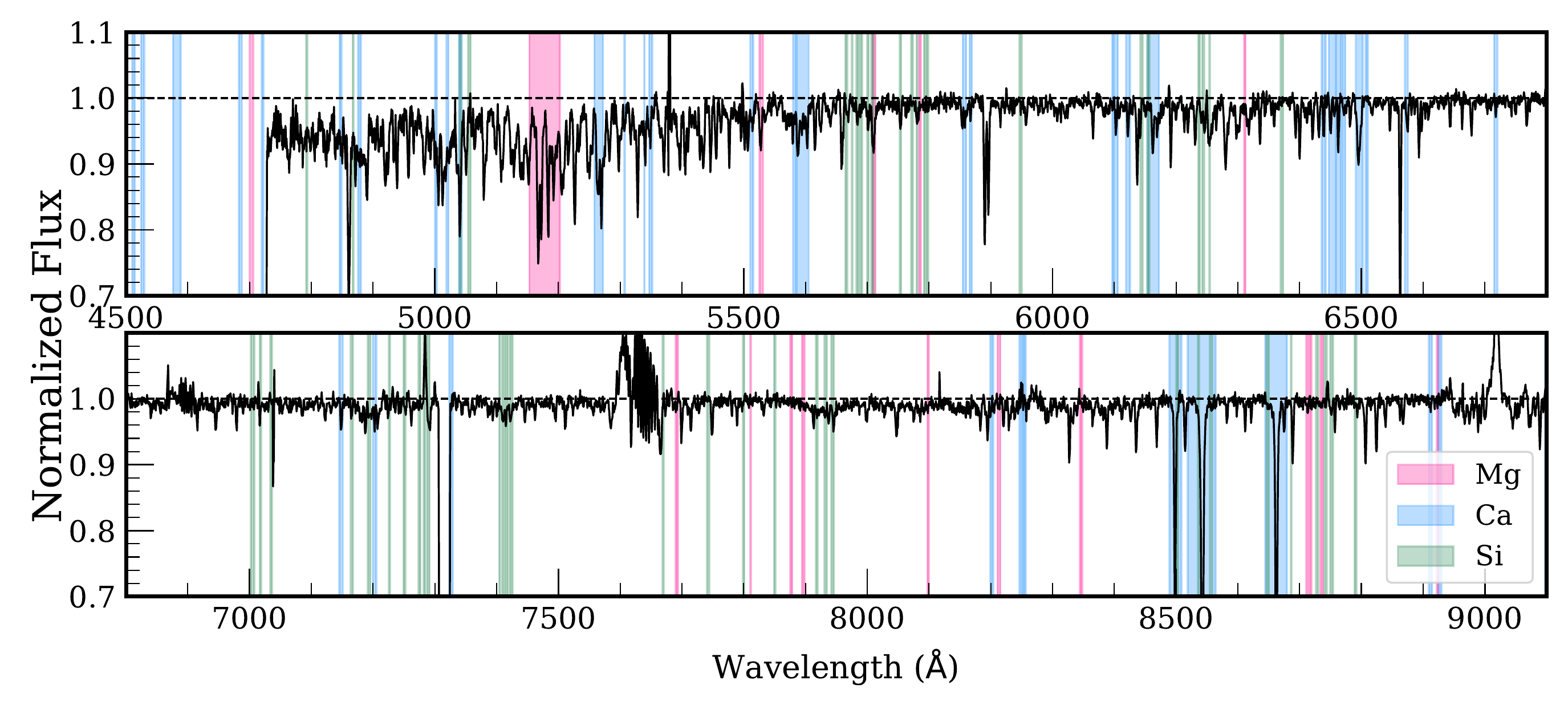}
\caption{Wavelength regions sensitive to changes in [$\alpha$/Fe] for the spectral resolution of the DEIMOS 600ZD grating ($\sim$ 2.8 \AA). We show an example spectrum (\textit{black}) over the wavelength range 4500$-$\deleted{6300}\added{9100} \AA, where we corrected the spectrum for telluric absorption (\S~\ref{sec:telluric}) and performed an initial continuum normalization (\S~\ref{sec:cont_norm}). We do not show spectral regions with wavelengths below 4500 \AA, since low S/N generally prevents utilization of the observed spectrum in this wavelength range. The spectrum is for a star in the globular cluster NGC 2419. Spectral regions sensitive to Mg, Ca, and Si are shown as highlighted ranges in magenta, blue, and green respectively. The atmospheric value of [$\alpha$/Fe] is measured using the union of the Mg, Ca, and Si spectral regions.
\label{fig:spec_regions_alpha}}
\end{figure*}

We employ the ATLAS9 \citep{1993PhST...47..110K} grid of model stellar atmospheres, with no convective overshooting \citep{Castelli1997NotesAtmospheres}. We base our grid on recomputed (\citealt{Kirby2009} and references therein) ATLAS9 model atmospheres with updated opacity distribution functions, available for [$\alpha$/Fe] = 0.0 and +0.4 \citep{CastelliKurucz2004}. We adopt the solar composition of \citealt{AndersGrevesse1989}, except for Fe \citep{Sneden1992}. The elements considered to be $\alpha$-elements are O, Ne, Mg, Si, S, Ar, Ca, and Ti. 

\begin{table}
\centering
\caption{Parameter Ranges of Blue Grid (4100 - 6300 \AA)}
\label{tab:grid}
\begin{tabular}{lccc}
\hline
\hline
Parameter & Minimum Value & Maximum Value & Step\\
\hline
$T_{eff}$ (K) & 3500 & 5600 & 100\\
& 5600 & 8000 & 200\\
$\log$ $g$ (cm s$^{-2}$) & 0.0 \ ($T_{eff}$ $<$ 7000 K) & 5.0 & 0.5\\
& 0.5 \ ($T_{eff}$ $>$ 7000 K) & 5.0 & 0.5\\
\big[Fe/H\big] & $-$4.5\footnote{Below [Fe/H] $<$ $-$4.5 for T$_{eff}$ $\leq$ 4100 K, certain stellar atmosphere models fail to converge when solving for molecular equilibrium in each atmospheric layer. Synthetic spectra with [Fe/H] $<$ $-$4.5 exist for a majority of T$_{eff}$-$\log$ $g$ pairs for T$_{eff}$ $\leq$ 4100 K, but our grid is complete for all parameter combinations only above [Fe/H] = $-$4.5 in this regime.} \ ($T_{eff}$ $\leq$ 4100 K) & 0.0 & 0.1\\
& $-$5 \ ($T_{eff}$ $>$ 4100 K) & 0.0 & 0.1\\
\big[$\alpha$/Fe\big] & $-$0.8 & 1.2 & 0.1\\
\hline
\end{tabular}
\end{table}

For stellar parameters between grid points, we linearly interpolated to generate model atmospheres within the ranges 3500 K $<$ T$_{eff}$ $<$ 8000 K, 0.0 $<$ $\log$ $g$ $<$ 5.0, $-$4.5 $<$ [Fe/H] $<$ 0.0, and -0.8 $<$ [$\alpha$/Fe] $<$ +1.2. A full description of the grid is presented in Table~\ref{tab:grid}.
%Here, [A/H] represents the \textit{total} metallicity of the spectrum, augmenting the abundances of all elements except H and He.
Here, [$\alpha$/Fe] represents a \textit{total} $\alpha$-element abundance for the atmosphere, which augments the abundances of individual $\alpha$-elements without distinguishing between their relative abundances. In total, the grid contains 316,848 synthetic spectra. 
%Assuming a total [A/H] and [$\alpha$/H] enables measurements of the abundances of individual $\alpha$- and Fe-peak elements using the same model atmosphere, breaking the consistency between atmospheric and spectral abundances.

We generated the synthetic spectra using MOOG \citep{1973PhDT.......180S}, an LTE spectral synthesis software. MOOG takes into account neutral hydrogen collisional line broadening \citep{Barklem2000ACollisions,Barklem2005TheCollisions}, in addition to radiative and Stark broadening and van der Waals line damping. The most recent version (2017) includes an improved treatment of Rayleigh scattering in the continuum opacity (Alex Ji, private communication). The resolution of each generated synthetic spectrum is 0.02 \AA.

\subsection{Photometric Constraints}
\label{sec:phot}

To reduce the dimensionality of parameter space and to optimize our ability to find the global chi-squared minimum in the parameter estimation (\S~\ref{sec:spec_abund}), we constrained the effective temperature and surface gravity of the synthetic spectra by available \added{Johnson-Cousins $VI$} photometry for red giant stars in our sample. The photometric effective temperature is estimated using a combination of the Padova \citep{Girardi2002}, Victoria-Regina \citep{VandenBerg2006}, and Yonsei-Yale \citep{Demarque2004} sets of isochrones, assuming an age of 14 Gyr and an $\alpha$-element abundance of 0.3 dex. If available, we also employed the \citet{RamirezMelendez2005} color temperature. We adopted a single effective temperature ($T_\textrm{eff,phot}$) and associated uncertainty ($\sigma_{T_\textrm{eff,phot}}$) from an average of the isochrone/color temperatures for each star. 

%We calculate the average of the different effective temperatures from the aforementioned sources, weighted by the inverse square of the uncertainties in the photometric temperature. We estimate the error on the photometric effective temperature by the quadratic sum of the weighted standard deviation of the weighted mean and the mean of the errors, weighted by the inverse square of the uncertainties. This calculated average photometric effective temperature and its uncertainty is used to constrain the spectroscopic effective temperature in the chi-squared minimization procedure (\S~\ref{sec:spec_abund}).

We determined the photometric surface gravity in a similar fashion. However, no color-$\log$ $g$ relation exists, so we could not include this additional source for the photometric surface gravity. Unlike the effective temperature, we did not solve for $\log$ $g$ using spectral synthesis techniques, as the errors on the photometric surface gravity are negligible when the distance is known. Additionally, LRS and MRS spectra cannot effectively provide constraints on its value owing to the lack of ionized lines in the spectra. Thus, we held $\log g$ fixed in the abundance determination.

\subsection{[Fe/H] and [$\alpha$/Fe] Regions}
\label{sec:spec_regions}

\begin{figure*}
\centering
\includegraphics[width=\textwidth]{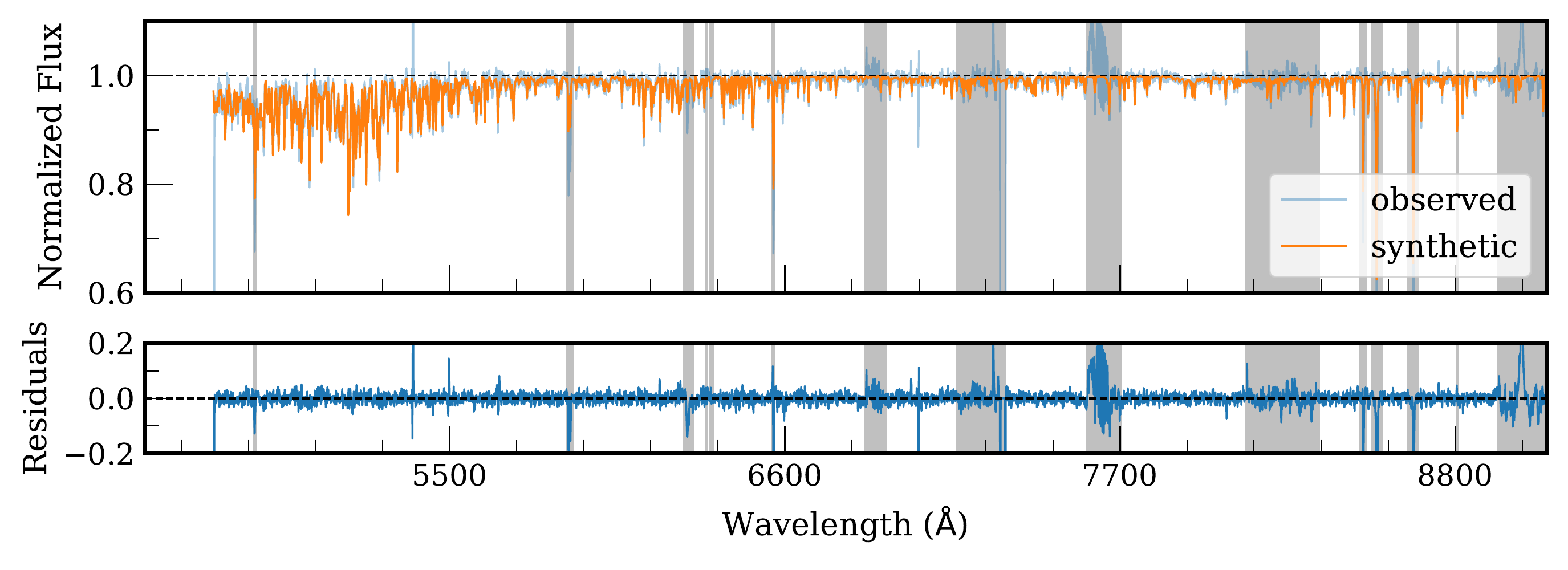}
\caption{An example of a continuum-normalized observed spectrum (\textit{light blue}) and its best-fit synthetic spectrum (\textit{orange}). The observed spectrum corresponds to the same object in Figure~\ref{fig:spec_regions_alpha}. We show the \deleted{portion of the fit utilizing our new blue grid of synthetic spectra ($\sim$ 4700 - 6300 $\AA$)} \added{the entire fitted wavelength range ($\sim$ 4500 - 9100 \AA) for this object}. The highlighted regions (\textit{grey}) correspond to our standard mask (\S~\ref{sec:custom_mask}), which excludes lines such as H$\beta$, Na D1 and D2 from the fit. We adopt the parameters of the best-fit synthetic spectrum for the observed spectrum, $T_\textrm{eff}$ = \deleted{4296}\added{4300} K, $\log$ $g$ = 0.71 dex, [Fe/H] = \deleted{$-$1.98}\added{$-$2.07} dex, [$\alpha$/Fe] = \deleted{0.18}\added{0.29} dex, and $\Delta\lambda$ = \deleted{2.66}\added{2.60} \AA \ FWHM. We measure $\chi_\nu^2$ = \deleted{1.64}\added{1.63} for the quality of the fit across the full wavelength range \deleted{($\sim$ 4700 - 9100 \AA)}, based on the regions of the spectrum used to measure [Fe/H] (\S~\ref{sec:spec_regions}). The normalized residuals (\textit{dark blue}) between the continuum-refined observed spectrum and best-fit synthetic spectrum are also shown. The residuals have been scaled by the inverse variance of the observed spectrum and the degree of freedom of the fit, such that each pixel represents the direct contribution to $\chi^2_\nu$.}
\label{fig:spec_abund}
\end{figure*}

In order to increase the sensitivity of the synthetic spectrum fit to a given abundance measurement, we constructed wavelength masks that highlight regions that are particularly responsive to changes in [Fe/H] and [$\alpha$/Fe]. We employed the same procedure as \citet{Kirby2009} to make the masks, starting with a base synthetic spectrum for each combination of $T_\textrm{eff}$ (3500 - 8000 K in steps of 500 K) and $\log$ g (0.0 - 3.5 in steps of 0.5 dex). We assumed a bulk metallicity [Fe/H] = -1.5 and solar [$\alpha$/Fe] for the atmosphere. We then generated synthetic spectra with either enhanced or depleted values of individual element abundances (Fe, Mg, Si, Ca, and Ti) for each $T_{\rm{eff}}$-$\log$ $g$ pair and compare to the base synthetic spectra, identifying wavelength regions that differ by more than 0.5\%. In the determination of the [Fe/H] and [$\alpha$/Fe] wavelength regions, we smoothed all synthetic spectra used to an approximation of the expected resolution of the 600ZD grating ($\sim$ 2.8 \AA) across the entire spectrum (4100 - 9100 \AA). We then compared the spectral regions for each element against the line list and high signal-to-noise (S/N $>$ 100) spectra of cool ($T_{\rm{eff}}$ $<$ 4200 K) globular cluster stars, eliminating any regions that do not have a corresponding transition in the line list or an absorption feature in the spectra.

Although our measurements reflect the atmospheric value of [$\alpha$/Fe], we constructed the associated wavelength mask from the regions sensitive to changes in the individual elements Mg, Si, and Ca. We excluded Ti from the [$\alpha$/Fe] mask owing to the prevalence of regions sensitive to Ti at bluer optical wavelengths, such that we cannot meaningfully isolate its elemental abundance. Figure~\ref{fig:spec_regions_alpha} illustrates our [$\alpha$/Fe] mask \deleted{in the blue} \added{across the entire usable spectral range} (4500$-$\deleted{6300}\added{9100} \AA; \added{\S~\ref{sec:obs}}). The [Fe/H] spectral regions cover 92\% and 51\% of the wavelength range in the red and blue respectively, whereas the [$\alpha$/Fe] regions span 15\% and 12\% of the same wavelength ranges. The overlap between the [Fe/H] regions and [$\alpha$/Fe] regions is 16\% and 23\% in the blue and red respectively. We emphasize that [Fe/H] and [$\alpha$/Fe] are measured separately and iteratively (\S~\ref{sec:spec_abund}).

\subsection{Parameter Determination from Spectral Synthesis}
\label{sec:spec_abund}

Here, we outline the steps involved in our measurement of atmospheric parameters and elemental abundances from spectral synthesis of low-resolution spectra. Figure~\ref{fig:spec_abund} illustrates an example of a final continuum-normalized observed spectrum and best-fit synthetic spectrum resulting from our measurement procedure for a high S/N RGB star in a MW GC.
Our method is nearly identical to that of \citet{Kirby2009}, excepting our introduction of an additional free parameter, $\Delta\lambda$, the resolution of the observed spectrum. We use a Levenberg-Marquardt algorithm to perform each comparison between a given observed spectrum and a synthetic spectrum. We weight the comparison according to the inverse variance of the observed spectrum. In each step, the synthetic spectra utilized in the minimization are interpolated onto the observed wavelength array and smoothed to the fitted observed resolution, $\Delta\lambda$, prior to comparison with a given observed spectrum.

\begin{enumerate}
\item \textit{$T_\textrm{eff}$, [Fe/H], and $\Delta\lambda$, first pass.} All three parameters are allowed to vary simultaneously in the fit. We use only regions sensitive to [Fe/H] (\S~\ref{sec:spec_regions}) in this measurement. We choose to measure $\Delta\lambda$ simultaneously with $T_\textrm{eff}$ and [Fe/H] to prevent the chi-squared minimizer from under- or over-smoothing the synthetic spectrum to compensate for the initial guesses of $T_\textrm{eff}$ and [Fe/H], which are offset from the final parameter values corresponding to the global $\chi^2$ minimum. The [Fe/H] regions cover almost the entire spectrum (92\%) in the wavelength range 4100 - 6300 \AA \ and a majority of the spectrum (51\%) in the range 6300 - 9100 \AA, such that using the entire spectrum to measure $\Delta\lambda$ does not change the results within the statistical uncertainties.

We assume a starting value of $T_{\textrm{eff,phot}}$ (\S~\ref{sec:phot}) for the spectroscopic effective temperature. $T_{\textrm{eff}}$ is constrained by photometry using a Gaussian prior, such that $\chi^2$ increases if $T_\textrm{eff}$ deviates substantially from $T_{\textrm{eff,phot}}$, as defined by the associated error in the photometric effective temperature ($\sigma_{T_{\rm{eff,phot}}}$). As motivated in \S~\ref{sec:phot}, $\log$ $g$ is fixed at the photometric value in all steps. We initialize [Fe/H] at $-$2 dex, where we performed tests to ensure that the final value of [Fe/H] does not depend on the initial guess. Similar to the approach for $T_\textrm{eff}$, we enforce a Gaussian prior with a mean of 2.8 \AA \ and standard deviation of 0.05 \AA \ on $\Delta\lambda$, according to the expected spectral resolution for the 600ZD grating.\added{\footnote{Approximating the spectral resolution by constant value of $\Delta\lambda$ over the full spectrum does not impact the determination of either [Fe/H] or [$\alpha$/Fe]. Even if $\Delta\lambda$ over-smooths (under-smooths) the spectrum in the fitting procedure, it should not alter the identified $\chi^2$ minimum for [Fe/H] and [$\alpha$/Fe], given that the effect of over-smoothing (under-smoothing) the core and wings of an absorption feature should effectively negate each other. However, approximating the spectral resolution by $\Delta\lambda$ could increase $\chi^2_\nu$, consequently increasing the statistical uncertainties on the abundances.}} \added{For the first iteration of the continuum refinement,} [$\alpha$/Fe] remains fixed at solar\added{. In subsequent iterations, [$\alpha$/Fe] is fixed at the value determined in step 2.} \deleted{whereas} The other parameters are allowed to vary until the best-fit synthetic spectrum is found. 

\item \textit{[$\alpha$/Fe], first pass.} $T_\textrm{eff}$, [Fe/H], and $\Delta\lambda$ are fixed at the values determined in step 1 while [$\alpha$/Fe] is allowed to vary, assuming a starting value of solar \added{for the first iteration of the continuum refinement. Otherwise, the starting value is the value of [$\alpha$/Fe] determined in the last iteration of the continuum refinement. As in the case of [Fe/H], the final value of [$\alpha$/Fe] does not depend on the initial guess}. In the determination of the best-fit synthetic spectrum, only wavelength ranges sensitive to variations in $\alpha$-element abundance are considered (\S~\ref{sec:spec_regions}).

\item \textit{Iterative continuum refinement.} After a best-fit synthetic spectrum is determined according to steps 1 and 2, we refine the continuum normalization according to \S~\ref{sec:cont_norm}. We perform the continuum refinement iteratively, enforcing the convergence conditions that the difference in parameter values between the previous and current iteration cannot exceed 1 K, 0.001 dex, 0.001 dex, and 0.001 \AA \ for $T_\textrm{eff}$, [Fe/H], [$\alpha$/Fe], and $\Delta\lambda$ respectively. If these conditions are not met in a given iteration, the continuum-refined spectrum is used to repeat steps 1 and 2 until convergence is achieved. If the maximum number of iterations (N$_{\textrm{iter, max}}$ = 50) is exceeded, which occurs for a small fraction of observed spectra, we do not include the observed spectra in the subsequent analysis.

\item \textit{[Fe/H], second pass.} [Fe/H] is redetermined, where $T_\textrm{eff}$\added{,} \deleted{and}$\Delta\lambda$\added{, and [$\alpha$/Fe]} are fixed at their converged values from step 3\deleted{ and [$\alpha$/Fe] = 0}. We use the final continuum-refined observed spectrum determined in step 3 in this step and all remaining steps.

\item \textit{[$\alpha$/Fe], second and final pass.} We repeat step 2, holding [Fe/H] fixed at the value determined in step 4.

\item \textit{[Fe/H], third and final pass.} We repeat step 4, holding [$\alpha$/Fe] fixed at the value determined in step 5.
\end{enumerate}

\begin{figure}
\centering
\includegraphics[width=\columnwidth]{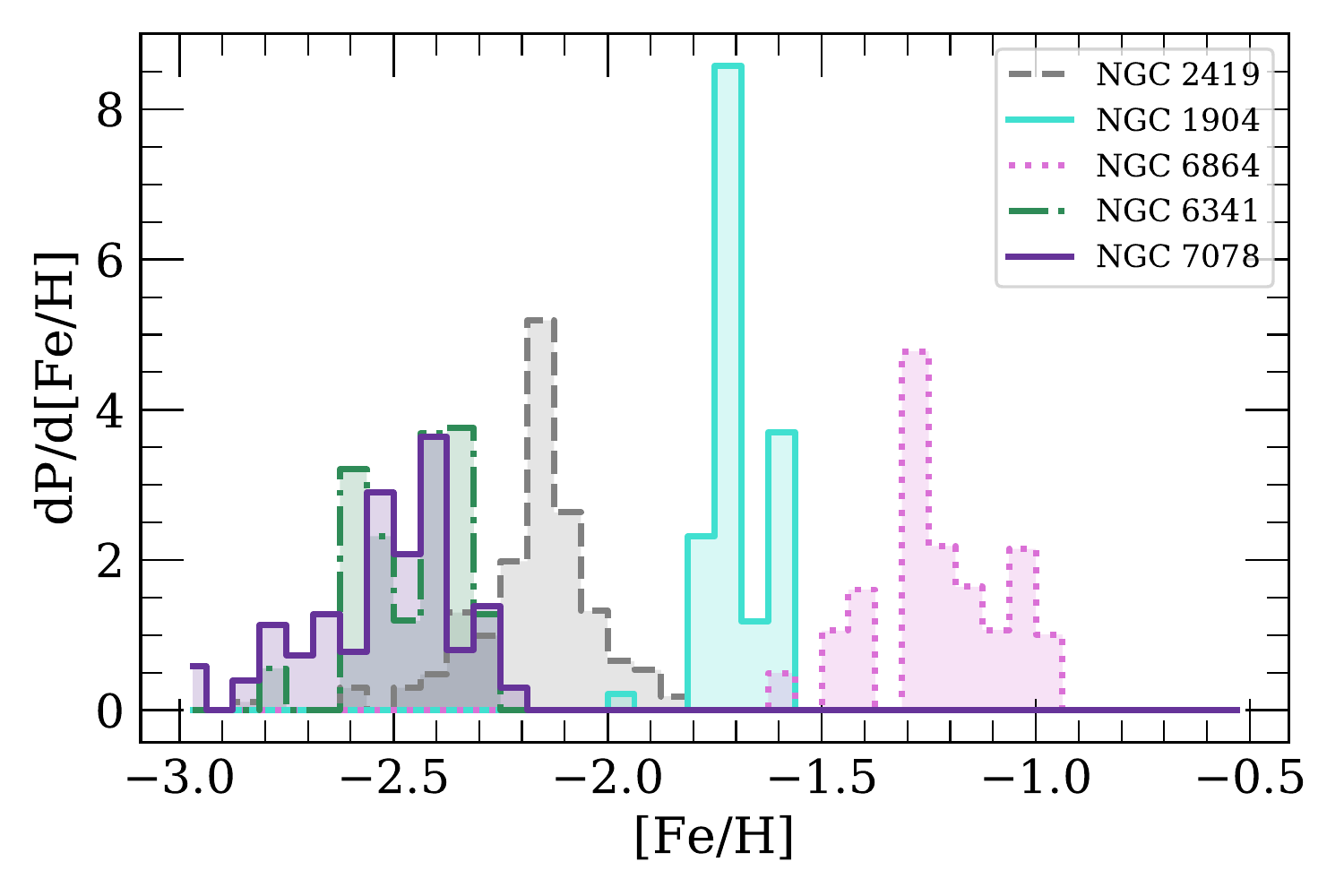}
\caption{Error-weighted metallicity ([Fe/H]) distribution functions for RGB members of Galactic globular clusters NGC 2419 (\textit{grey}), NGC 1904 (\textit{cyan}), NGC 6864 (\textit{magenta}), NGC 6341 (\textit{green}), \added{and NGC 7078 (\textit{purple})}. Only stars for which $\delta$[Fe/H] $<$ 0.3 dex are shown. We find mean cluster metallicites of \deleted{$-$2.08 dex, $-$1.63 dex, $-$1.19 dex, and $-$2.45 dex}\added{$-$2.14 dex, $-$1.70 dex, $-$1.22 dex, $-2.45$ dex, and $-$2.50 dex} for the four respective clusters.}
\label{fig:mdf}
\end{figure}

%We then fix the spectral resolution at the average resolution for all objects in a given slitmask, and repeat the above procedure with a constant value of $\Delta\lambda$ (as opposed to allowing it to vary as a free parameter).

\section{Globular Cluster Validation Tests}
\label{sec:results}

We demonstrate the robustness of our LRS technique by applying it to a set of MW GCs: NGC 2419, NGC 1904 (M79), NGC 6864 (M75), and NGC 6341 (M92) \added{and NGC 7078 (M15)}. 

NGC 2419 is a luminous outer halo GC located $\sim$ 90 kpc away from the Galactic center \citep{Harris1997} with multiple stellar populations, but no detected variation in [Fe/H] \citep{CohenKirby2012}. NGC 6864 also exhibits evidence for chemically distinct populations, including a marginal spread in [Fe/H] ($\sim$ 0.07 dex; \citealt{Kacharov2013}).  It is a relatively young GC \citep{Catelan2002} at a Galactocentric radius of $\sim$ 15 kpc \citep{Harris1997}. NGC 1904 ($\sim$ 19 kpc; \citealt{Harris1997}) possesses an extended blue horizontal branch, but it is otherwise a typical cluster. NGC 6341 ($\sim$ 9 kpc; \citealt{Harris1997}) is notable primarily for being very metal-poor ([Fe/H] $\sim$ $-$2.3 dex). \added{NGC 7078 is similarly metal-poor, and has been observed to exhibit variations in $\alpha$-elements \citep{Sneden2000,Carretta2009a}}. For a summary of observations used in our validation tests, see Table~\ref{tab:obs}.

%We demonstrate the effectiveness and robustness via comparisons to previously existing abundance measurements from spectral synthesis of MRS (6300 - 9100 \AA) \citep{Kirby2008,Kirby2009} and various HRS data sets \citep{CohenKirby2012,Kacharov2013,Carretta2009a,Carretta2010}. The primary difference between the LRS and MRS methods is that we measure abundances across a broader wavelength range ($\sim$ 4500 - 9100 \AA). Additional nuances in differences in methodology are thoroughly addressed in \S~\ref{sec:prep_spec} and ~\ref{sec:abund}. We perform a detailed comparison to the various HRS approaches in \S~\ref{sec:hrs}. 
%We focus on the MW globular cluster (GC) NGC 2419 \citep{Harris1997}, for which we have abundances from LRS, MRS, and HRS. For a summary of observations used in our validation tests, see Table~\ref{tab:obs}.

\subsection{Membership}
\label{sec:membership}

For all subsequent analysis in this section, we utilize only stars that have been identified as RGB \added{or AGB} star members by \citet{Kirby2016}. \deleted{We removed asymptotic giant branch stars from our GC sample that \citet{Kirby2016} manually selected from color-magnitude diagrams.} Membership is defined using both radial velocity and metallicity criteria based on MRS, such that any star whose measurement uncertainties are greater than 3$\sigma$ from the mean of either radial velocity or metallicity is not considered a member. The colors and magnitudes of member stars must also conform to the cluster's giant branch.

\subsection{Metallicity}
\label{sec:gc_metallicity}

As described in \S~\ref{sec:spec_abund}, we measure metallicity from spectral regions sensitive to variations in [Fe/H]. In addition to membership criteria (\S~\ref{sec:membership}), we further refine our sample by requiring that the 5$\sigma$ contours in \deleted{each of the four fitted parameters}\added{$T_\textrm{eff}$, [Fe/H], and [$\alpha$/Fe]} (\S~\ref{sec:spec_abund}) identify the minimum. This condition is effectively equivalent to requiring that a given star has sufficient S/N, a converged continuum iteration, and overall high enough quality fit ($\chi^2_\nu$) to produce a reliable abundance measurement.

%We obtained spectra for a total of 93 stars in NGC 2419 (Table~\ref{tab:obs}), where we have photometry (\S~\ref{sec:phot}) and previous measurements from the red side data for 80 stars. We refine our sample by requiring that the continuum refinement process converges within the maximum allowable number of iterations (73 stars), $\chi^2_{\nu_\textrm{[Fe/H]}}$  $\leq$ 2 (65 stars), and S/N $>$ 20 (64 stars), resulting in 53 stars in our final sample.

\begin{figure}
\centering
\includegraphics[width=\columnwidth]{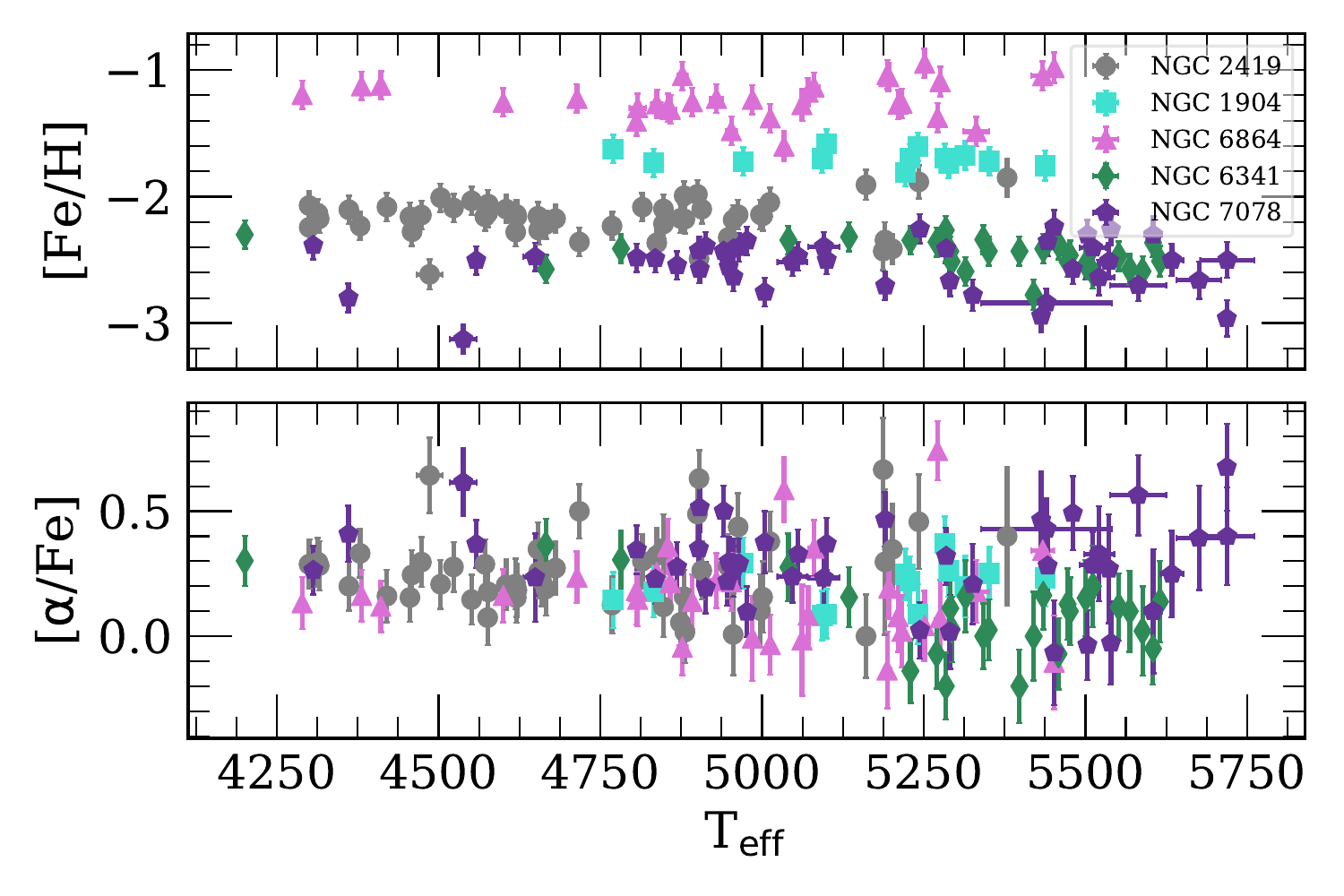}
\caption{\added{(\textit{Top})} metallicity ([Fe/H]) \added{and (\textit{bottom}) [$\alpha$/Fe]} vs. spectroscopic effective temperature ($T_{\textrm{eff}}$). \added{We show only stars with $\delta$[Fe/H] $<$ 0.3 dex and $\delta$[$\alpha$/Fe] $<$ 0.3 dex in each panel.}
%We show results from both LRS (this work; \textit{blue+red}) and MRS (\citealt{Kirby2008}; \textit{red}). We show total error for [Fe/H] and statistical uncertainty for $T_{\textrm{eff}}$.
The lack of a trend between \deleted{the two quantities} \added{both [Fe/H] and [$\alpha$/Fe] with respect to $T_{\textrm{eff}}$} for each GC implies that our chemical abundance analysis is robust to systematic covariance in these parameters. %\label{fig:teff_vs_feh_red_vs_blue}
\label{fig:teff_vs_feh}}
\end{figure}

We illustrate our results for [Fe/H] in the form of metallicity distribution functions (Figure~\ref{fig:mdf}) for NGC 2419, NGC 1904, NGC 6864, and NGC 6341, where we weight the distribution according to the \textit{total} error in [Fe/H]. For a discussion of the measurement uncertainties, including systematic uncertainties, see \S~\ref{sec:sys}. We find that $\langle$[Fe/H]$\rangle$ = $-$\deleted{2.08}\added{2.18} $\pm$ \deleted{0.21}\added{0.15} dex, $-$\deleted{1.63}\added{1.70} $\pm$ \deleted{0.07}\added{0.08} dex, $-$\deleted{1.19}\added{1.23} $\pm$ 0.15 dex, $-$2.45 $\pm$ \deleted{0.19}\added{0.12} dex, \added{and $-$2.53 $\pm$ 0.19 dex} for NGC 2419, NGC 1904, NGC 6864, NGC 6341, \added{and NGC 7078} respectively, where $\langle$[Fe/H]$\rangle$ is weighted according to the inverse variance of the total measurement uncertainty. These values \added{approximately} agree with the corresponding quantities from HRS: $-$2.12 $\pm$ 0.09 \citep{CohenKirby2012}, $-$1.58 $\pm$ 0.03 \citep{Carretta2009a}, $-$1.16 $\pm$ 0.07 \citep{Kacharov2013}, $-$2.34 dex \citep{Sneden2000}\footnote{\citet{Sneden2000} do not cite random uncertainties on their abundances. We represent their [Fe/H] and [$\alpha$/Fe] (\S~\ref{sec:hrs}) values as simple means.}, \added{and $-$2.32 $\pm$ 0.07 dex} respectively. \added{In particular, we note that we find a spread in [Fe/H] for M15 that is likely not intrinsic \citep{Carretta2009c}, but rather a consequence of measurement uncertainty. Our estimate of the systematic uncertainty in [Fe/H] (\S~\ref{sec:sys}) incorporates this dispersion in [Fe/H] measurements}. We present a detailed comparison \added{of our [Fe/H] measurements} to HRS abundances in \S~\ref{sec:hrs}. 

%We show our results for [Fe/H] from LRS (\textit{blue+red}) versus [Fe/H] from MRS (\textit{red}) in Figure~\ref{fig:mdf}, where the errors reflect the total error. The majority of data points fall on the one-to-one line for abundance measurements from LRS and MRS, indicating agreement between methods. We find that $\langle$[Fe/H]$\rangle_{\rm{LRS}}$ = -2.09 $\pm$ 0.14 dex and $\langle$[Fe/H]$\rangle_{\rm{MRS}}$ = -2.16 $\pm$ 0.09 dex, where $\langle$[Fe/H]$\rangle$ is weighted according to the inverse variance of the total measurement uncertainty. Although the MDF for the LRS measurements is skewed toward slightly higher metallicity and exhibits larger dispersion, the LRS and MRS measurements clearly agree within the uncertainties.

As another example of our ability to reliably recover \deleted{[Fe/H]} \added{abundances}, we show [Fe/H] \added{and [$\alpha$/Fe]} vs.\ spectroscopically determined $T_{\textrm{eff}}$ in Figure~\ref{fig:teff_vs_feh} for all GCs. In a nearly mono-metallic population like a GC, correlation of metallicity with other fitted parameters, such as $T_{\textrm{eff}}$, would indicate the presence of systematic effects. Because $T_{\rm{eff}}$ is strongly covariant with [Fe/H], the fitting procedure might erroneously select a lower value of [Fe/H] and $T_{\rm{eff}}$ in order to match spectral features. Figure~\ref{fig:teff_vs_feh} presents evidence against any such correlation. \added{The same argument can be extended to the $\alpha$-element abundance of a GC, given the assumption of chemical homogeneity. Similarly, we do not see any correlation between [$\alpha$/Fe] and $T_{\rm{eff}}$}.

\subsection{$\alpha$-element Abundance}

\begin{figure}
\centering
\includegraphics[width=\columnwidth]{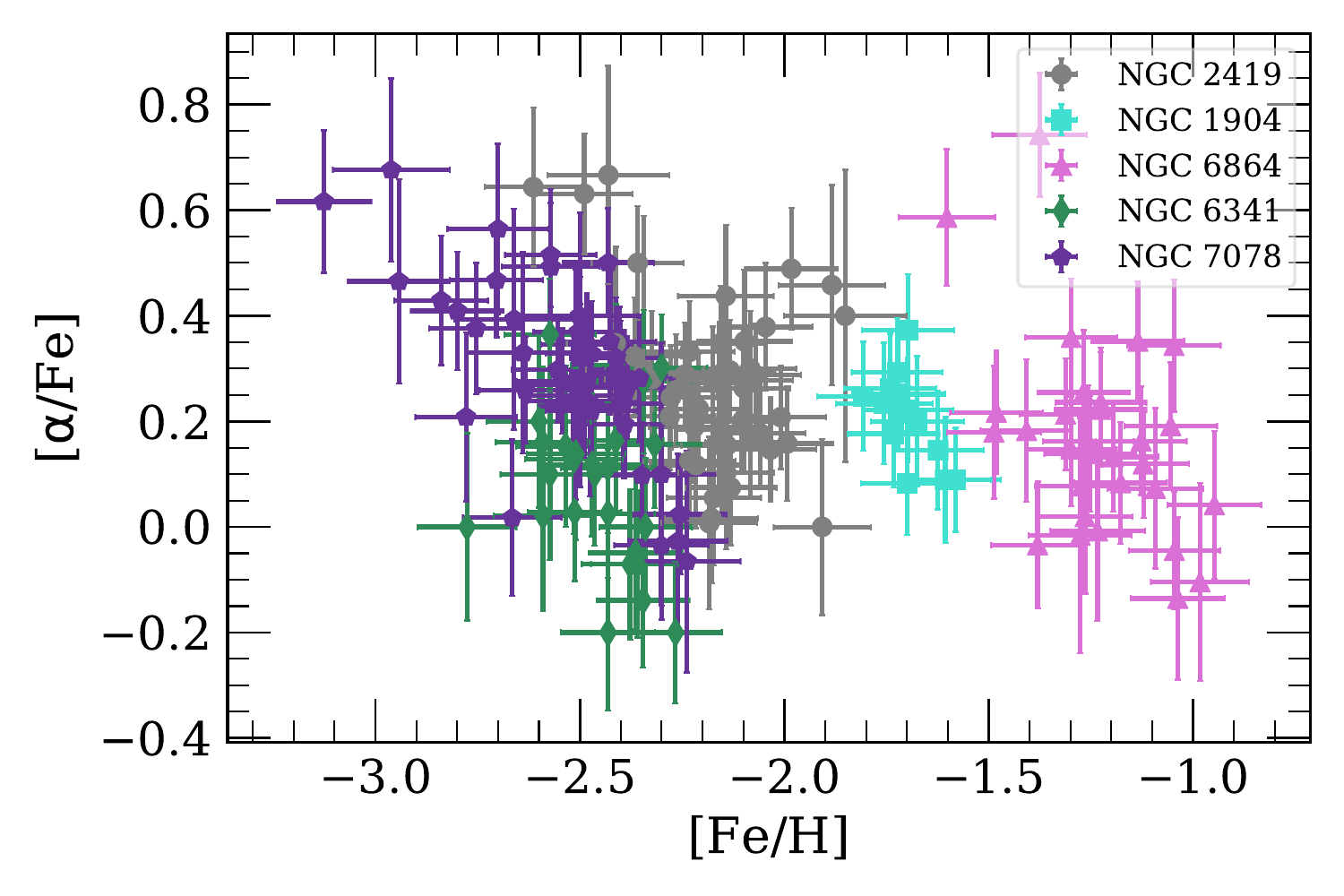}
\caption{Atmospheric [$\alpha$/Fe] versus [Fe/H] for the same data set as Figure~\ref{fig:mdf}, where we also exclude points with $\delta$([$\alpha$/Fe]) > 0.3 dex. There is no apparent anticorrelation between [$\alpha$/Fe] and [Fe/H] within a GC, indicating that our method does not show any unphysical covariance between these two parameters. \label{fig:alphafe_vs_feh}}
\end{figure}

Similarly, we do not anticipate a correlation between [Fe/H] and [$\alpha$/Fe] within a GC\@. [$\alpha$/Fe] abundance impacts the determination of [Fe/H] via its contribution of H$^{-}$ opacity to the stellar atmosphere through electron donation. Thus, the abundance of [$\alpha$/Fe] alters stellar atmospheric structure, requiring a re-evaluation of [Fe/H] in the spectral fitting process. The presence of trends between [$\alpha$/Fe] with [Fe/H] (e.g., increasing [$\alpha$/Fe] with decreasing [Fe/H]) within a GC, which we expect to contain no such correlations, would indicate systematic effects in measuring abundances. As summarized in Figure~\ref{fig:alphafe_vs_feh}, no such systematics are \added{consistently} present in our data for each GC. \added{Even in the worst case scenario of M15, a massive, very metal-poor GC with known $\alpha$-element variation, any apparent anticorrelation is primarily driven by a few outliers in both [Fe/H] and [$\alpha$/Fe].}

%In addition, our LRS measurements overlap with the [$\alpha$/Fe] distribution from MRS, although the LRS measurements show a smaller dispersion in [$\alpha$/Fe], where fewer stars have high [$\alpha$/Fe]. The smaller spread in the LRS measurements could be explained by the presence of additional $\alpha$-element absorption lines at blue wavelengths, such as the Mg I triplet (5167 \AA, 5173 \AA, and 5184 \AA).

\section{Comparison to High-Resolution Spectroscopy}
\label{sec:hrs}

\begin{figure}
\centering
\includegraphics[width=\columnwidth]{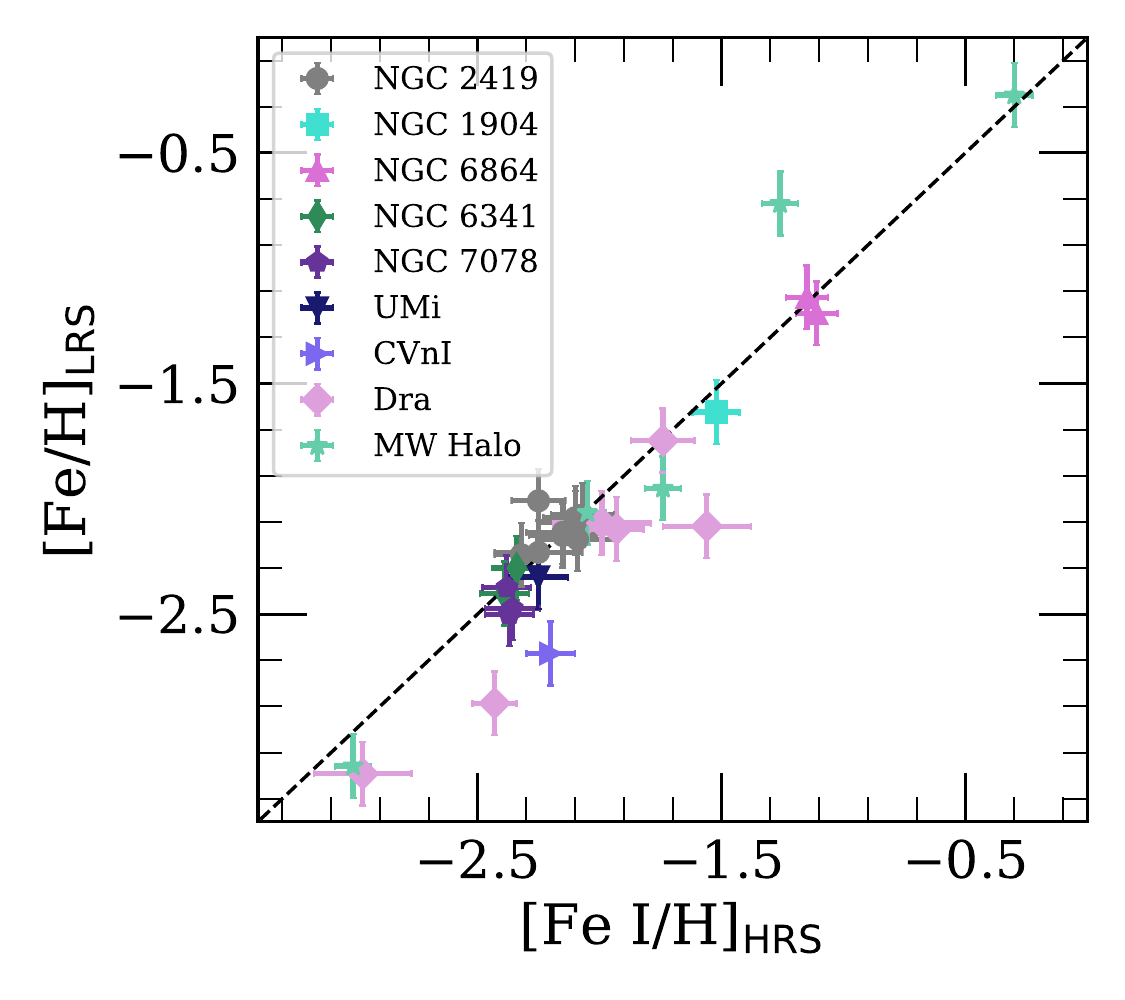}
\caption{Metallicity measured from LRS ([Fe/H]) versus HRS ([\ion{Fe}{1}/H]) \added{for MW GCs, dSphs, and halo stars.} \deleted{\citep{CohenKirby2012,Carretta2009a,Carretta2010,Kacharov2013,}. for NGC 2419, \added{NGC 1904,} and NGC 6864.}
Between LRS and HRS data sets, we have measurements in common for \deleted{9}\added{30} stars. Although some scatter is present between data sets, \deleted{they exhibit broad agreement} \added{[\ion{Fe}{1}/H]$_\textrm{HRS}$ and [Fe/H] are strongly correlated at high significance across multiple orders of magnitude in metallicity.} \deleted{within the 1$\sigma$ uncertainties ($\delta\epsilon_\textrm{sys}=0.02$ dex, Eq.~\ref{eq:hrs_comp}). No correlation exists between the LRS and HRS measurements, which we expect for a monometallic GC.} \label{fig:feh_hires}} 
\end{figure}

\subsection{High-Resolution Data}
\label{sec:hrs_data}

Given the variety in approaches of HRS studies of the MW GCs, \added{MW dSphs, and MW halo stars} listed in Table~\ref{tab:obs}, we provide a summary of the stellar parameter determination and abundance analysis in each case. For all \deleted{GCs}\added{systems}, membership is determined based on radial velocities. 

\begin{itemize}

    \item \textit{NGC 2419}:  Using Keck/HIRES ($R$ $\sim$ 34,000) spectroscopy, \citet{CohenKirby2012} measured [Fe/H], [Mg/Fe], [Si/Fe], and [Ca/Fe] for 13 RGB stars in NGC 2419. They used MOOG \citep{1973PhDT.......180S} in combination with \citet{CastelliKurucz2004} atmospheric models to derive equivalent widths from neutral lines across the wavelength range 4500 - 8350 \AA, including the Mg triplet.  Stellar parameters were set to photometric values. Measurement uncertainties represent the dispersion of the mean abundance based on the various lines used in the abundance determination.
    
    \item \textit{NGC 6864}: \citet{Kacharov2013} used Magellan/MIKE ($R$ $\sim$ 30,000) to observe 16 RGB stars in NGC 6864 over a wavelength range of 3340 - 9150 \AA. They measured [Fe/H], [Mg/H], [Si/H], and [Ca/H] via equivalent width measurements using MOOG and \citet{CastelliKurucz2004} atmospheric models. Mg was measured from a single line (5711 \AA). They determined T$_\textrm{eff}$ from excitation equilibrium and surface gravities from T$_\textrm{eff}$, extinction-corrected bolometric magnitude ($M_\textrm{bol}$), and the known distance to the cluster. The measurement uncertainties are a combination of the random error (based on the number of lines used in the abundance analysis for a given element) and a component that reflects the error from adopted stellar atmosphere parameters. For the latter component, we adopt the larger, more conservative errors that reflect averages based on the entire GC sample of \citet{Kacharov2013}. 
    
    \added{\item \textit{NGC 1904}: From VLT/UVES ($R$ $\sim$ 40,000), \citet{Carretta2009a} and \citet{Carretta2010} performed an abundance analysis based on equivalent width measurements for Fe, Mg, Si, and Ca, respectively, for a sample of 10 RGB stars, over the wavelenth range 4800 - 6800 \AA. Following \citet{Carretta2009b}, the authors adopted $T_\textrm{eff}$ from calibrated $V$-$K$ colors and surface gravity from $T_\textrm{eff}$, $M_\textrm{bol}$, and distance moduli, and used Kurucz atmosphere models (with convective overshooting). To determine errors in the stellar atmospehric parameters, \citeauthor{Carretta2009a}\ repeated their analysis for each star, varying a single atmospheric parameter each time, to derive an average internal error, in addition to the rms error.}
    
    \added{\item \textit{NGC 6341}: Based on WIYN/Hydra (R $\sim$ 20,000, $5740$ \AA $< \lambda\lambda < 5980$ \AA spectra, \citet{Sneden2000} measured [Fe/H], [Si/Fe], and [Ca/Fe] abundances for RGB stars in NGC 6341 and NGC 7078. A single Si transition is used to determine [Si/Fe]. They adopted atmospheric parameters from $B-V$ photometry calibrations and employed MARCS atmosphere models in combination with MOOG. Given that \citet{Sneden2000} did not provide an estimate of abundance errors, we assume an uncertainty of 0.1 dex for all elemental abundances.}
    
    \added{\item \textit{NGC 7078}: We utilize a compilation of data from \citet{Carretta2009a,Carretta2010} and \citet{Sneden1997,Sneden2000}. The latter two studies employed similar techniques, where \citet{Sneden1997} used a mix of observations from both Hamilton and HIRES. In contrast to \citet{Sneden2000}, \citet{Sneden1997} measured Mg abundances, based on a combination of equivalent width measurements $1-2$ strong \ion{Mg}{1} lines and spectral synthesis of a slightly weaker line. We use [\ion{Fe}{1}/H], as opposed to [Fe/H], which is given as the mean of [\ion{Fe}{1}/H] and [\ion{Fe}{2}/H] and assume 0.1 dex uncertainties on the HRS abundances.}
    
    \added{\item \textit{MW dSphs}: For Canes Venatici I (CVnI) and Ursa Minor (UMi), we find a single star %SDSS J132755.65+333324.5
    in common between each of our DEIMOS slitmasks and the HRS literature \citep{Francois2016, Shetrone2001}.  \citet{Francois2016} used VLT/X-shooter spectra ($\lambda$ = 300 nm $-$ 1 $\mu$m, $R = 7900-12600$) to measure [Fe/H], [Mg/Fe], and [Ca/Fe] for two stars in CVnI\@. They determined $T_\textrm{eff}$ using a color-temperature relation and $V, I_C$ photometry, $\log g$ from $M_\textrm{bol}$, and abundances via spectral synthesis using OSMARCS \citep{Gustafsson1975, Plez1992} atmosphere models. We estimated the HRS abundance errors based on the typical uncertainties of the published data. Based on Keck/HIRES spectroscopy ($4540$ \AA $ \lesssim \lambda\lambda \lesssim 7020 $\AA ), \citet{Shetrone2001} measured [Fe/H], [Mg/Fe], and [Ca/Fe] for several stars in UMi, using MOOG and MARCS model atmospheres. They provided large upper limits on [Si/Fe], which we exclude from our analysis. Atmospheric parmaters were determined simultaneously and iteratively using dereddened $(B-V)$ color-temperature relations, excitation equilibrium, and ionization balance. As for Draco, we use a mixture of data \citep{Shetrone1998, Shetrone2001, Fulbright2004, CohenHuang2009}. The methods of \citet{Shetrone1998} are nearly identical to those of \citet{Shetrone2001}, as is the case for \citet{CohenHuang2009} in relation to \citet{CohenKirby2012}. Aside from using Keck/HIRES spectroscopy and adopting photometric values for atmospheric parameters, the analysis of \citet{Fulbright2004} is similar to that of \citet{Fulbright2000} (see discussion of MW halo HRS abundances). Additionally, the majority of Draco stars do not have a HRS measurement of Si abundance.}
    
    \added{\item \textit{MW Halo}: We selected 5 MW halo stars from \citet{Fulbright2000} with $\log g$ $<$ 3.5 dex and [$\alpha$/Fe] $\lesssim$ 0.15 dex or [$\alpha$/Fe] $\gtrsim$ 0.4 dex. Using high-resolution ($R$ $\sim$ 50,000), high S/N Lick/Hamilton, \citet{Fulbright2000} measured Mg, Si, and Ca for 168 halo and disk stars. $T_\textrm{eff}$ and $\log$ $g$ were determined iteratively using Fe lines, with initial guesses determined from $V-K$ photometry. Abundances were determined from equivalent width measurements using MOOG and Kurucz model atmospheres with convective overshooting.  We use [\ion{Fe}{1}/H], as opposed to [Fe/H], which is given as the mean of [\ion{Fe}{1}/H] and [\ion{Fe}{2}/H]. The abundance errors reflect the rms uncertainty in each elemental abundance.}
    
\end{itemize}

We emphasize that the HRS abundances do not include true estimates of systematic uncertainty, e.g., resulting from limitations from the selected grid of model atmospheres or line list. Additionally, we are comparing our homogenous LRS abundances to an inhomogenous HRS sample.  As a result, some of the differences among the HRS studies can be attributed simply to different abundance measurement tools and techniques.

\subsection{Abundance Comparison}
\label{sec:hrs_results}

\begin{figure}
\centering
\includegraphics[width=\columnwidth]{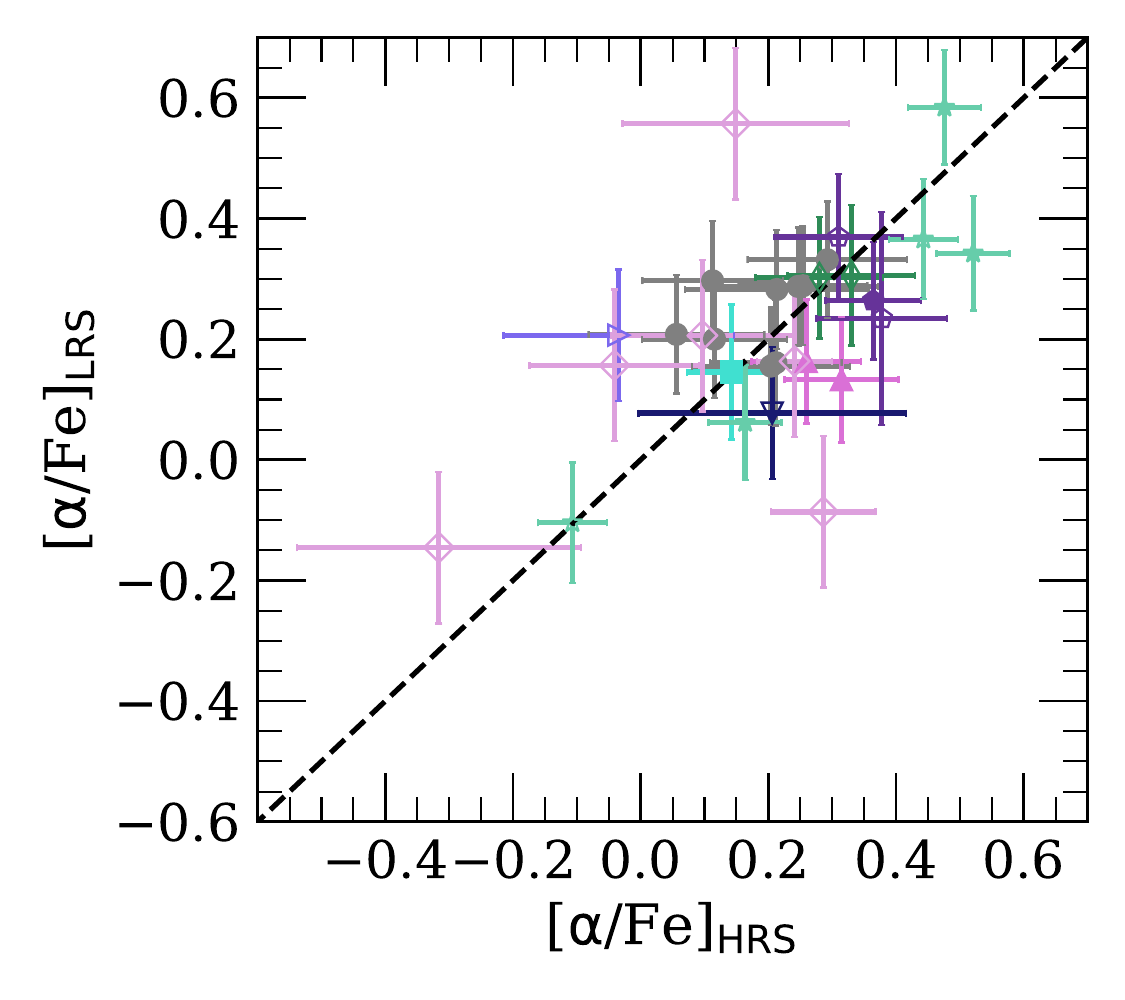}
\caption{[$\alpha$/Fe]$_\textrm{HRS}$ versus [$\alpha$/Fe]$_\textrm{LRS}$ for \deleted{NGC 2419 and NGC 6864}\added{the same data set as Fig.~\ref{fig:feh_hires}}. We construct [$\alpha$/Fe]$_\textrm{HRS}$ based on a weighting (\S~\ref{sec:hrs_results}) of its individual $\alpha$-element abundances. \added{Open symbols indicate values of [$\alpha$/Fe]$_\textrm{HRS}$ constructred from two or fewer individual $\alpha$-element abundance ratios.} The LRS and HRS abundances exhibit broad agreement \deleted{($\delta\epsilon_\textrm{sys}$ consistent with zero, Eq.~\ref{eq:hrs_comp})}\added{ and show a clear correlation}, despite the intrinsic differences between the sets of measurements.
\label{fig:alphafe_hires}} 
\end{figure}

%Performing our abundance measurement over a similarly broad wavelength range (4500 - 9100 \AA)
We find \deleted{reasonable}\added{good} agreement in [Fe/H] for the \deleted{9}\added{30} stars common between both data sets, \added{where we have included abundance measurements of MW dSphs and MW halo stars, in addition to GCs, to expand our sample size with HRS overlap.}.\deleted{ within the 1$\sigma$ uncertainties.} In order to perform the comparison, we have shifted the HRS abundances \citep{CohenMelendez2005a,Gratton2003,Asplund2009,GrevesseSauval1998, AndersGrevesse1989} to the same solar abundance scale as the LRS abundances. Figure~\ref{fig:feh_hires} shows \deleted{no correlation between the LRS and HRS metallicity measurements within each of NGC 2419 and NGC 1904, which is expected for monometallic globular clusters}\added{a strong correlation between the LRS and HRS measurements across a wide metallicity range ($\sim -3.0 - 0.0$ dex)}.

In order to compare our [$\alpha$/Fe] measurements to an analogous HRS quantity, we construct [$\alpha$/Fe]$_\textrm{HRS}$ based on a weighted sum of Mg, Si, and Ca elemental abundances. To derive the weights, we start with a reference synthethic spectrum defined by T$_\textrm{eff}$ = 4400 K, $\log$ $g$ = 1.0 dex, [Fe/H] = $-$1.8 dex, which correspond to mean parameter values from HRS studies of NGC 2419, NGC 1904, NGC 6864, and NGC 6341, and [$\alpha$/Fe] = 0. We assume a spectral resolution of $\Delta\lambda$ = 2.8 \AA \ and interpolate the synthetic spectrum onto a wavelength array with spacing equal to the pixel scale of the 600 line mm$^{-1}$ grating ($\sim$ 0.64 \AA). Next, we enhance/deplete the $\alpha$-element abundance by 0.1, 0.2, and 0.3 dex, calculating the sum of the absolute difference between the reference and enhanced/depleted synthetic spectrum in each case. For each $\alpha$-element, we utilize only the relevant wavelength regions (\S~\ref{sec:spec_regions}) and spectral coverage that corresponds to our data set (4500 - 9100 \AA). Additionally, we exclude contributions from masked wavelength regions (\S~\ref{sec:custom_mask}). We adopt the normalized average value of the summed absolute flux differences as our final weight for a given element, i.e.,

\begin{equation}
\label{eq:alphafe_hires}
\begin{split}
[\alpha/\textrm{Fe}]_\textrm{HRS} = 0.282 & \times \textrm{[Mg/Fe]}_\textrm{HRS} + 0.136 \times \textrm{[Si/Fe]}_\textrm{HRS} \\
& + 0.582 \times \textrm{[Ca/Fe]}_\textrm{HRS},
\end{split}
\end{equation}

\begin{equation}
\label{eq:alphafeerr_hires}
\begin{split}
\delta [\alpha/\textrm{Fe}]_\textrm{HRS} = \Bigl[ \bigl( 0.282 & \times \delta\textrm{[Mg/Fe]}_\textrm{HRS} \bigr)^2 + \bigl( 0.136 \times \delta\textrm{[Si/Fe]}_\textrm{HRS} \bigr)^2 \\
& + \bigl( 0.582 \times \delta\textrm{[Ca/Fe]}_\textrm{HRS} \bigr)^2 \Bigr]^{1/2}.
\end{split}
\end{equation}
%\right \right & \left \times \delta\textrm{[Mg/Fe]}_\textrm{HRS} \right)^2 + \bigl( 0.136 \times \delta\textrm{[Si/Fe]}_\textrm{HRS} \bigr)^2 \\
%& \left + \left( 0.582 \times \delta\textrm{[Ca/Fe]}_\textrm{HRS} \right)^2 \right]^{1/2}.

In Figure~\ref{fig:alphafe_hires}, we utilize Eqs.~\ref{eq:alphafe_hires} and ~\ref{eq:alphafeerr_hires} to directly compare HRS and LRS $\alpha$-element abundances \deleted{in NGC 2419 and NGC 6864} \added{for the same sample as Figure~\ref{fig:feh_hires}}. \added{A clear positive correlation exists between [$\alpha$/Fe]$_\textrm{LRS}$ and [$\alpha$/Fe]$_\textrm{HRS}$, with some degree of scatter present.} We emphasize that [$\alpha$/Fe]$_\textrm{HRS}$ represents only an approximation to the atmospheric value of [$\alpha$/Fe], given the fundamental differences between the HRS and LRS methods (\S~\ref{sec:hrs_data}).

\begin{table}
\centering
\begin{threeparttable}
\caption{$\langle$[$\alpha$/Fe]$\rangle$ in MW GCs}
\begin{tabular}{lcc}
\hline\hline
GC & $\langle$[$\alpha$/Fe]$_\textrm{LRS}\rangle$ (dex) & $\langle$[$\alpha$/Fe]$_\textrm{HRS}\rangle$ (dex) \\
\hline
NGC 2419 & \added{0.26} $\pm$ \added{0.13} & 0.21 $\pm$ 0.09 \\
NGC 1904 & \added{0.20} $\pm$ \added{0.09} & 0.28 $\pm$ 0.02 \\
NGC 6864 & \added{0.18} $\pm$ \added{0.18} & 0.28 $\pm$ 0.07 \\
NGC 6341 & \added{0.10} $\pm$ \added{0.16} & 0.37 \\
NGC 7078 & 0.30 $\pm$ 0.15 & 0.32 $\pm$ 0.06 \\
\hline
\end{tabular}
\begin{tablenotes}
\item Note.\textemdash \ The HRS references for NGC 2419, NGC 1904, NGC 6864, NGC 6341, \added{and NGC 7078} are \citet{CohenKirby2012}, \citet{Carretta2009a,Carretta2010}, \citet{Kacharov2013}, and \citet{Sneden2000}. \added{We construct $\langle$[$\alpha$/Fe]$_\textrm{HRS}\rangle$ from a weighting of the available individual HRS $\alpha$-element abundances (Eq.~\ref{eq:alphafe_hires}). No published HRS Mg measurements exist for NGC 6341, whereas $\langle$[$\alpha$/Fe]$_\textrm{LRS}\rangle$ includes Mg in all cases.}
\end{tablenotes}
\label{tab:alphafe_hrs_gc_comp}
\end{threeparttable}
\end{table}

In Table~\ref{tab:alphafe_hrs_gc_comp}, we summarize our findings for $\langle$[$\alpha$/Fe]$\rangle_\textrm{LRS}$ in MW GCs and compare to equivalent HRS measurements constructed using Eqs.~\ref{eq:alphafe_hires} and ~\ref{eq:alphafeerr_hires}.\footnote{The HRS reference used to construct [$\alpha$/Fe]$_\textrm{HRS}$ for NGC 7078 is based on the most recent study of $\alpha$-enhancement in this cluster. The star-to-star HRS comparisons of Figures~\ref{fig:feh_hires} and \ref{fig:alphafe_hires} are based on a compilation of values from \citet{Sneden1997, Sneden2000} and \citet{Carretta2009a,Carretta2010} for NGC 7078.} In the case of NGC 6341, we construct [$\alpha$/Fe]$_\textrm{HRS}$ based only on [Ca/Fe] and [Si/Fe] because \citet{Sneden2000} did not measure [Mg/Fe]. \added{In any instance of an incomplete set of Mg, Ca, and Si abundances for our HRS comparisons,} we renormalized the weights in Eqs.~\ref{eq:alphafe_hires} and ~\ref{eq:alphafeerr_hires} accordingly. The average LRS $\alpha$-element abundances are \added{in some cases} lower than the HRS measurements by $\sim$0.1 dex, \added{but overlap within the associated standard deviation on the measurement}, excepting the case of NGC 6341. We find a significantly lower value of $\langle$[$\alpha$/Fe]$\rangle$, although we note that it is particularly difficult to compare between LRS and HRS in this case. 
Given that NGC 6341 lacks \added{HRS} Mg abundances, \added{we cannot perform a meaningful direct comparison of our constructed $\langle$[$\alpha$/Fe]$_\textrm{HRS}\rangle$ to our measurements of $\langle$[$\alpha$/Fe]$_\textrm{LRS}\rangle$, which include Mg.}

%We find that $\langle$[$\alpha$/Fe]$\rangle_\textrm{LRS}$ = 0.13 $\pm$ 0.09, 0.11 $\pm$ 0.05, and 0.13 $\pm$ 0.11, and 0.07 $\pm$ 0.13 dex for the full samples of NGC 2419, NGC 1904, NGC 6864, and NGC 6341 respectively. Using Eqs.~\ref{eq:alphafe_hires} and ~\ref{eq:alphafeerr_hires}, the equivalent HRS measurements are $\langle$[$\alpha$/Fe]$\rangle_\textrm{HRS}$ = 0.21 $\pm$ 0.09 \citep{CohenKirby2012}, 0.28 $\pm$ 0.02 \citep{Carretta2009a, Carretta2010}, 0.28 $\pm$ 0.07 \citep{Kacharov2013}, and 0.37 dex \citep{Sneden2000}\footnote{We construct [$\alpha$/Fe]$_\textrm{HRS}$ based only on [Ca/Fe] and [Si/Fe] because \citet{Sneden2000} did not measure [Mg/Fe].  We renormalized the weights in Eqs.~\ref{eq:alphafe_hires} and ~\ref{eq:alphafeerr_hires} accordingly.} 

\deleted{Despite} This apparent offset in \deleted{the cluster means} \added{between the LRS and HRS measurements can be characterized in terms of a systematic uncertainty component}. We find that a star-by-star comparison of [$\alpha$/Fe] and [Fe/H] shows that our LRS results are consistent with those from HRS within the \deleted{random}\added{systematic} uncertainties (\S~\ref{sec:sys}). We \deleted{prove this consistency}\added{estimate the systematic error} by calculating the additional error term that would be required to force the LRS and HRS measurement to agree within one standard deviation.  The relevant equation is

\begin{figure}
    \centering
    \includegraphics[width=\columnwidth]{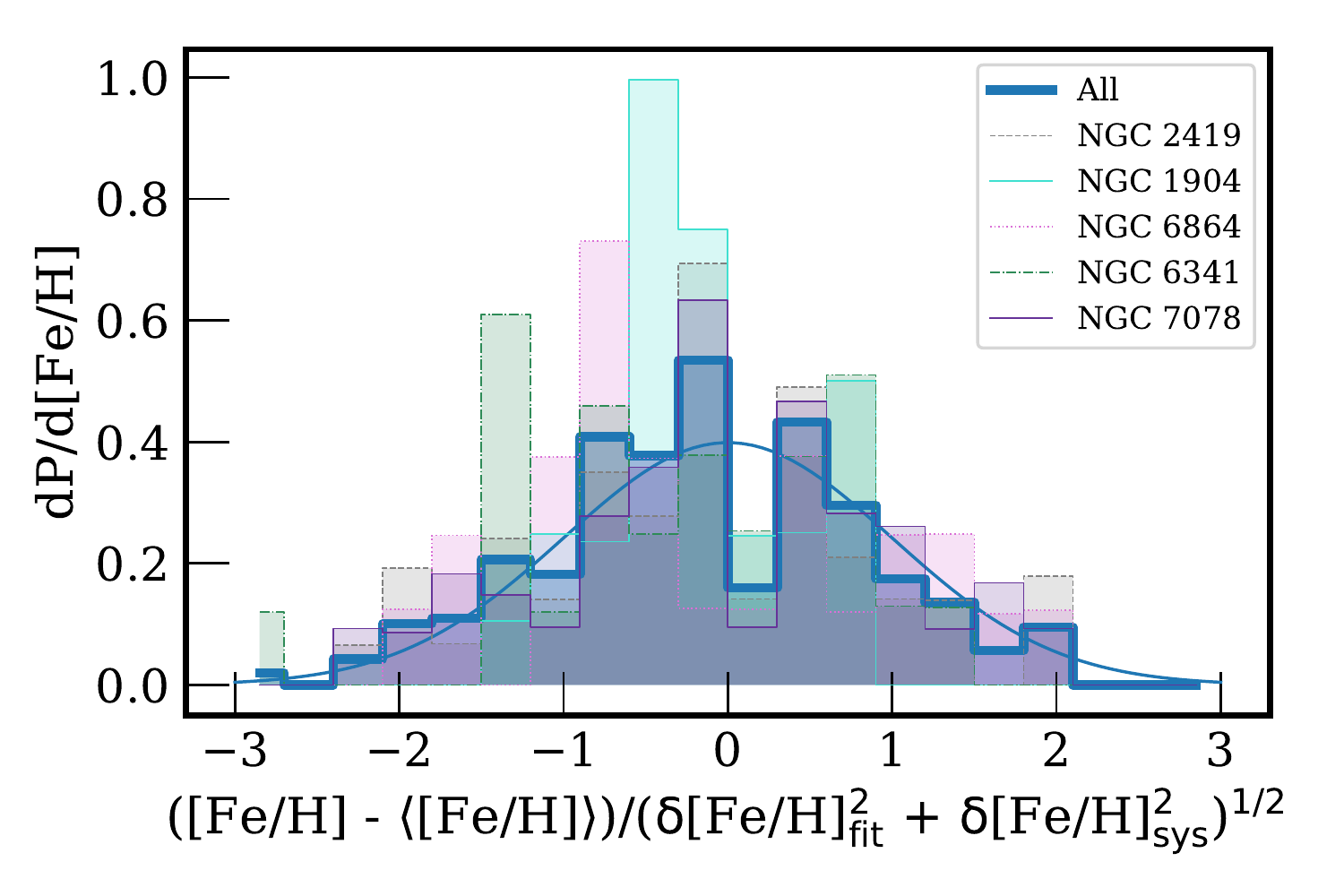}
    \caption{Probability distribution function of [Fe/H] normalized to the mean metallicity of a given cluster ($\langle$[Fe/H]$\rangle$) and weighted by the total error in metallicity. We show the distributions for NGC 2419 (\textit{grey}), NGC 1904 (\textit{cyan}), NGC 6864 (\textit{magenta}), NGC 6341 (\textit{green}) \added{NGC 7078 \textit{purple}}, and all \deleted{four}\added{five} clusters (\textit{blue}) (\deleted{117}\added{154} stars). The total error is composed of the statistical uncertainty from the fit ($\delta$[Fe/H]$_\textrm{fit}$) and the systematic uncertainty ($\delta$[Fe/H]$_\textrm{sys}$). We determine the systematic uncertainty from the intrinsic dispersion in the combined distribution for all three clusters. The Gaussian defined by the systematic uncertainty ($\delta$([Fe/H])$_\textrm{sys}$ = \deleted{0.105}\added{0.111} dex) is overplotted.}
    \label{fig:feherr_sys}
\end{figure}

\begin{equation}
\frac{1}{N} \sum_{i}^{N} \frac {\left( \epsilon_\textrm{LRS,i} - \epsilon_\textrm{HRS,i} \right)^2 } {\left( \delta\epsilon_\textrm{LRS,i}^2 + \delta\epsilon_\textrm{HRS,i}^2 + \delta\epsilon_{sys}^2 \right) } = 1,
\label{eq:hrs_comp}
\end{equation}
where $\epsilon$ represents a given elemental abundance, such as [Fe/H] or [$\alpha$/Fe], $\delta\epsilon$ is the corresponding statistical uncertainty on the measurement, $i$ is an index representing a given star in common between both the HRS and LRS data sets, and $N$ is the total number of common stars. \deleted{We do not find a solution for $\delta\epsilon_\textrm{sys}$ in the case of \added{both [Fe/H] and} [$\alpha$/Fe], indicating that, even without a systematic component, the measurements are consistent.}

\added{For $\epsilon$ = [Fe/H] and $N$ = 30 stars, we numerically solve Eq.~\ref{eq:hrs_comp} to find $\delta\epsilon_\textrm{sys}$ = 0.176 dex. A majority of this systematic term is driven by a single MW halo star with a discrepant value of [Fe/H]$_\textrm{LRS}$. Excluding it from the calculation, we find $\delta\epsilon_\textrm{sys}$ = 0.143 dex for $N$ = 29 stars. This value is likely more representative of the true systematic uncertainty, given that it is not subject to extreme outliers and exhibits better agreement with an independent calculation of the systematic uncertainty from the intrinsic spread in GCs (\S~\ref{sec:sys}, Table~\ref{tab:sys}). In the case of $\epsilon$ = [$\alpha$/Fe] we find $\delta\epsilon_\textrm{sys}$ = 0.058 dex and 0.039 dex, respectively, in the cases of $N$ = 30 and $N$ = 29. For consistency, we adopt the latter value to reflect the systematic uncertainty in [$\alpha$/Fe]$_\textrm{LRS}$ from HRS comparisons. Given the intrinsic heterogeneity of the HRS sample (\S~\ref{sec:hrs_data}) and the comparatively limited sample size ($N$ = 29 versus $N=154$), we chose not to adopt these values as our systematic uncertainty. Instead, we favor values calculated from the internal spread in GCs (\S~\ref{sec:sys}).}

\deleted{For the case of [Fe/H], we find $\delta\epsilon_\textrm{sys}$ = 0.02, which is negligible.} \deleted{Thus, upon inclusion of our systematic uncertainties determined from intrinsic dispersion in GCs (\S~\ref{sec:sys}), [Fe/H]$_\textrm{LRS}$ and [$\alpha$/Fe]$_\textrm{HRS}$ agree with the HRS data within 1$\sigma$.}

\section{Systematic Uncertainty from Internal Spread in Globular Clusters}
\label{sec:sys}

The total uncertainty on fitted parameters is composed of two components added in quadrature, the statistical (fit) uncertainty, $\delta_\textrm{fit}$, and a systematic component, $\delta_\textrm{sys}$. The fit uncertainty is calculated according to the reduced chi-squared value ($\chi^2_\nu$) and the diagonals of the covariance matrix of the fit ($\sigma_{ii}$), i.e., $\sigma_{ii} (\chi^2_\nu)^{1/2}$. We calculate $\chi^2_\nu$ using only the regions of the observed spectrum utilized in the fit, e.g., in the case of [Fe/H], we use the wavelength regions sensitive to [Fe/H] (\S~\ref{sec:spec_regions}) and not excluded by the pixel mask (\S~\ref{sec:custom_mask}). The systematic component encapsulates uncertainty intrinsic to our method, owing to sources such as the linelist (\S~\ref{sec:line_list}), assumptions involved in spectral synthesis (\S~\ref{sec:synth_spec}), details of our method, such as the continuum normalization (\S~\ref{sec:cont_norm}) and fitting procedure (\S~\ref{sec:spec_abund}), and covariance with other fitted parameters.\footnote{We do not completely characterize the systematic uncertainty on T$_\textrm{eff}$ and $\Delta\lambda$ because our primary goal is to determine abundances. The systematic errors $\delta$([Fe/H])$_\textrm{sys}$ and $\delta$([$\alpha$/Fe])$_\textrm{sys}$ already account for errors propagated by inaccuracies in T$_\textrm{eff}$ and $\Delta\lambda$. All of the of uncertainties in T$_\textrm{eff}$ and $\Delta\lambda$ presented in this paper reflect only the statistical uncertainty.} 

\subsection{Metallicity}

\begin{table}
\centering
\caption{Systematic Uncertainty}
\label{tab:sys}
\begin{tabular}{lcccc}
%\begin{tabular*}{0.5\columnwidth}{l @{\extracolsep{\fill}} c}
\hline\hline
Parameter & $\delta_\textrm{sys,HRS}$ (dex) & $N_\textrm{HRS}$\footnote{Number of stars with both LRS and HRS abundance measurements used to determine the systematic uncertainty from HRS} & $\delta_\textrm{sys,gc}$\footnote{The systematic uncertainty as calculated from the intrinsic spread in GCs (\S~\ref{sec:sys}). We adopt these values over the systematic uncertainty determined from comparison to HRS, $\delta_\textrm{sys,HRS}$ (\S~\ref{sec:hrs_results}), given the heterogeneity of the HRS sample (\S~\ref{sec:hrs_data}).} (dex) & $N_\textrm{gc}$\footnote{Number of stars used to determine the systematic uncertainty from the intrinsic spread in GCs.} \\
\hline
\big[Fe/H\big] & 0.143 & 29 & \added{0.111} & \added{154} \\
\big[$\alpha$/Fe\big] & 0.039 & 29 & \added{0.094} & \added{68} \\
\hline
%\end{tabular*}
\end{tabular}
\end{table}

Because most GCs, including those in our sample, are nearly monometallic \citep{Carretta2009c}, we can derive an estimate of systematic uncertainty in [Fe/H], $\delta$[Fe/H]$_{\textrm{sys}}$, by enforcing the condition that the intrinsic dispersion in the GC is zero, i.e.,
\begin{equation}
\sigma^2 =  \textrm{var} \left[ \frac{\rm{[Fe/H]}_i - \langle\rm{[Fe/H]}\rangle}{ \left( \delta\rm{[Fe/H]}_{\rm{fit},i}^2 + \delta\rm{[Fe/H]}_{\rm{sys}}^2 \right)^{1/2}} \right] = 1,
\label{eq:feherrsys}
\end{equation}
where $i$ is the index for a star in the GC, $\delta$[Fe/H]$_{\textrm{fit}}$ is the S/N-dependent statistical uncertainty in [Fe/H], and $\langle$[Fe/H]$\rangle$ is the mean metallicity of the GC, where the mean is weighted by the statistical uncertainty on each measurement of [Fe/H]\@. Eq.~\ref{eq:feherrsys} follows a reduced chi-squared distribution with an expectation value of unity. Enforcing the condition $\sigma^2$ = 1, we can numerically solve for the most likely value of the systematic uncertainty.

First, we refine our sample by removing outliers in [Fe/H] for each GC. Following \citet{Kirby2016}, we calculate the mean metallicity, $\langle$[Fe/H]$\rangle$, and standard deviation, $\sigma$([Fe/H]), for each cluster, and we remove stars that deviate by more than 2.58$\sigma$ (99\% confidence level)\@. Then, we re-compute $\langle$[Fe/H]$\rangle$ and $\sigma$([Fe/H]) from this refined sample, including only stars from the full sample that fulfill the criteria $|{\rm [Fe/H]}-\langle{\rm [Fe/H]}\rangle|- \delta({\rm [Fe/H]})_\textrm{fit} < 3\sigma$.  The inclusion of $\delta({\rm [Fe/H]})_\textrm{fit}$ in the inequality allows stars to be considered members even if they fall outside of the allowed metallicity range, as long as some part of their $1\sigma$ confidence intervals falls within the range.

After performing this sigma clipping for each individual cluster, we subtract the mean cluster metallicity from each star's measurement of [Fe/H], and we solve for the intrinsic dispersion based on the combined sample (Eq.\ \ref{eq:feherrsys}). We obtain a systematic uncertainty in [Fe/H] of $\delta$([Fe/H])$_\textrm{sys}$ = \deleted{0.105}\added{0.111} dex (Table~\ref{tab:sys}) based on \deleted{117}\added{154} stars. We present an illustration of this method in Figure~\ref{fig:feherr_sys}, where we show the probability distributions for the total-error-weighted metallicity of each cluster, in addition to the the combined GC sample. The fact that the combined distribution is well-approximated by a Gaussian with $\sigma$ = 1 indicates that the calculated systematic uncertainty sufficiently accounts for the observed metallicity spread.

\begin{figure}
    \centering
    \includegraphics[width=\columnwidth]{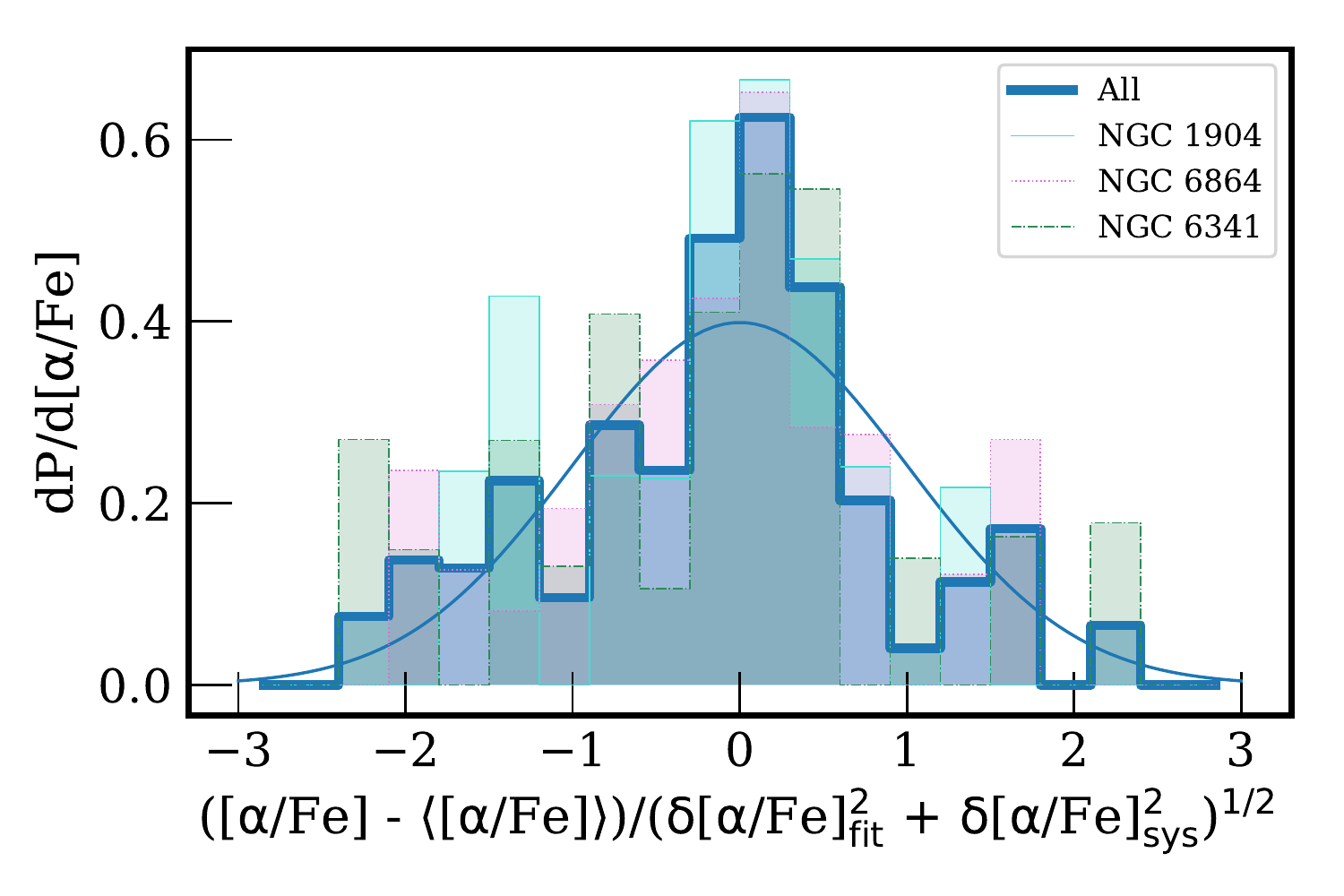}
    \caption{Probability distributions used to determine the systematic uncertainty, as in Figure~\ref{fig:feherr_sys}, except for the case of [$\alpha$/Fe]. We show the distributions for NGC 1904 (\textit{cyan}), NGC 6864 (\textit{magenta}), NGC 6341 (\textit{green}), and all three clusters (\textit{blue}) (\added{68 stars}). We find that $\delta$([$\alpha$/Fe])$_\textrm{sys}$ = \deleted{0.039}\added{0.094} dex. (\S~\ref{sec:alphafe_sys})}
    \label{fig:alphafeerr_sys}
\end{figure}

Thus, the total error is,

\begin{equation}
\delta\rm{([Fe/H])}_{\rm{tot}} = \sqrt{ \delta\rm{([Fe/H])}_{\rm{fit}}^2 + \delta\rm{([Fe/H])}_{sys}^2}
\label{eq:fehtoterr}
\end{equation}
In general, the statistical fit uncertainty \added{for [Fe/H]} is negligible compared to the systematic error for GCs. However, this \deleted{will}\added{is} not be the case for M31, given the low value of the expected S/N.

\subsection{$\alpha$-element Abundance}
\label{sec:alphafe_sys}

To determine the systematic uncertainty in [$\alpha$/Fe], $\delta$([$\alpha$/Fe])$_\textrm{sys}$, we calculate the intrinsic dispersion in the clusters, analogously to Eq.~\ref{eq:feherrsys}. Whereas it is generally reasonable to assume that GCs have negligible spread in [Fe/H], the assumption of zero intrinsic variation in [$\alpha$/Fe] must be evaluated individually for each cluster. \deleted{We exclude NGC 2419 from our combined GC sample in this case, given that}\added{For example,} abundance analysis of HRS has detected a significant spread in Mg \added{for NGC 2419, where a minority of the population is Mg-abnormal} \citep{Cohen2011, CohenKirby2012}. \added{Large star-to-star variations in Mg have also been found for NGC 7078 from HRS studies \citep{Sneden2000,Carretta2009a}}. Although NGC 6864 possesses chemically distinct populations, O is the only $\alpha$-element that exhibits significant variation within the cluster, as opposed to Mg, Si, or Ca \citep{Kacharov2013}. NGC 6341 is not known to possess $\alpha$-element variations \citep{Sneden2000}, with the caveat that no recent Mg abundances from HRS have been published to our knowledge. 

We therefore construct our combined GC sample from NGC 1904, NGC 6864, and NGC 6341 to compute $\delta$([$\alpha$/Fe])$_\textrm{sys}$, obtaining a value of \deleted{0.039}\added{0.094} dex (Table~\ref{tab:sys}) from \deleted{69}\added{68} stars. 
%\added{Including NGC 2419, we obtain $\delta$([$\alpha$/Fe])$_\textrm{sys}$ = 0.085 dex from 101 stars, in agreement with the value calculated by excluding NGC 2419. Given that $\delta$([$\alpha$/Fe])$_\textrm{sys}$ agrees between the two cases, we adopt the former value ($\delta$([$\alpha$/Fe])$_\textrm{sys}$ = 0.085 dex) which excludes NGC 2419.} 
Figure~\ref{fig:alphafeerr_sys} illustrates that the adopted error floor in [$\alpha$/Fe] describes the data well. We anticipate a smaller value of $\delta$([$\alpha$/Fe])$_\textrm{sys}$ relative to $\delta$([Fe/H])$_\textrm{sys}$, given that the systematic effects (uncertainties in the line list, atmospheric parameters, continuum normalization, etc.)\ that impact [Fe/H] tend to similarly affect [$\alpha$/H]. \deleted{Therefore, the net effect on the [$\alpha$/Fe] ratio is zero to first order.}

\section{The Star Formation History of the Stellar Halo of M31}
\label{sec:m31}

\begin{table}
\centering
\caption{M31 Stellar Halo Observations}
%\caption{M31 Stellar Halo Observations (f130\_2)}
\label{tab:m31_obs}
\begin{tabular}{lccccc}
\hline
\hline
Object & Date & $\theta_s$ ('') & $\langle X \rangle$ & $t_{\textrm{exp}}$ (s) & $N$\\
\hline
f130\_2a & 2018 Jul 19 & 1.0 & 1.53 & 5639 & 37\\
f130\_2b\footnote{Slitmasks indicated ``a'' and ``b'' are identical, except that the slits are titled according to the median parallactic angle at the approximate time of observation.} & 2018 Jul 19 & 1.0 & 1.16 & 5758 & 37\\
f130\_2a & 2018 Aug 14 & 0.86 & 1.29 & 4140 & 37\\
f130\_2a & 2018 Oct 10 & 0.83 & 1.84 & 3000 & 37\\
f130\_2a & 2018 Oct 11 & 0.60 & 1.49 & 2400 & 37\\
%f123\_1a & 2017 Oct 23 & 0.93 & 1.49 & 10800 & 104 & 1200G \\
\hline
\end{tabular}
\end{table}

We apply our spectral synthesis technique to spectra of individual RGB stars in the stellar halo of M31. We select a field with no identified substructure \added{\citep{Gilbert2007}} as an example.  We will apply our method to additional \added{M31} stellar halo fields in future work.

\subsection{Halo Field Observations}
\label{sec:m31_obs}

%\begin{figure*}
%    \centering
%    \includegraphics[width=\textwidth]{f130_2_mgt.pdf}
%    \caption{The Mg triplet region (5150$-$5200 \AA) for a star in f130\_2. We show the final continuum-normalized observed spectrum (S/N = 22.5 \AA$^{-1}$) (\textit{black}) and its best fit synthetic spectrum (\textit{red}), defined by T$_\textrm{eff}$ =  4356 $\pm$ 3 K, $\log$ $g$ = 0.78 dex, [Fe/H] = -2.19 $\pm$ 0.12 dex, [$\alpha$/Fe] = 0.43 $\pm$ 0.18 dex, and $\Delta\lambda$ = 2.82 $\pm$ 0.02 \AA. Holding all other parameters fixed, we modify the best-fit spectrum by $\delta$([$\alpha$/Fe]) = 0.18 dex (\textit{shaded red regions}) as an illustration of the uncertainty in [$\alpha$/Fe].}
%    \label{fig:f130_2_mgt}
%\end{figure*}

The field, f130\_2, is located at 23 kpc in projected radius along the minor axis of M31, and was first observed and characterized by \citet{Gilbert2007} using the Keck II/DEIMOS 1200 line mm$^{-1}$ grating. We selected it owing to its proximity to the 21 kpc halo field of \citet{Brown2007}, for which \citet{Brown2009} presented catalogs of deep optical photometry obtained using the Advanced Camera for Surveys (ACS) on the \textit{Hubble Space Telescope}. 

Table~\ref{tab:m31_obs} summarizes our observations of the M31 stellar halo field, which we observed with the same configuration as described in \S~\ref{sec:obs}.  The total exposure time was 5.8 hours. Following \citet{Cunningham2016}, we designed two separate slitmasks for the single field, with the same mask center, mask position angle, and target list, but with differing slit position angles. Switching slitmasks in the middle of the observation allows us to approximately track the change in parallactic angle over the course of the night. This technique mitigates flux losses due to differential atmospheric refraction (DAR), which disproportionately affects blue wavelengths. Thus, it is especially important to consider DAR when observing with the 600 line mm$^{1}$ grating, which covers a wider spectral range than any other DEIMOS grating.

\subsection{Sample Selection}
\label{sec:m31_sample}

\begin{figure}
    \centering
    \includegraphics[width=\columnwidth]{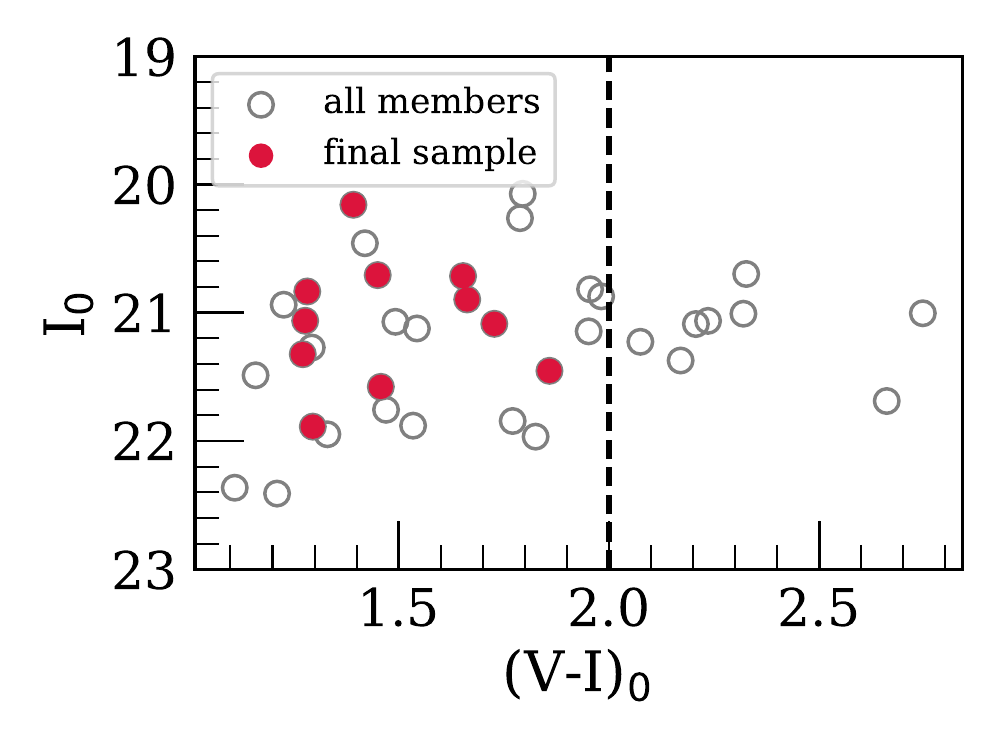}
    \caption{\added{($I$, $V-I$) color-magnitude diagram for M31 RGB stars from a 23 kpc field (f130\_2) with no identified substructure. We show both stars contained in the final sample (\textit{red filled circles}) (\S~\ref{sec:m31_sample}) and the full sample of member stars (\textit{grey open circles}).
    The dashed vertical line represents a rough threshold in color above which stars are likely to show evidence for strong TiO absorption in their spectra ($(V-I)_0$ $>$ 2.0). The final sample of stars shows significant overlap with the full distribution of M31 members, aside from known color biases that we have introduced in our sample selection.}}
    \label{fig:cmd}
\end{figure}

The observed field, at a M31 galactocentric radius of 23 kpc, includes a non-negligible contamination fraction of Milky Way foreground dwarf stars. In order to identify \deleted{secure}\added{both secure and marginal} M31 members \added{($\langle L_{i,\textrm{no v}} \rangle > 0$; \citealt{Gilbert2012})}, we used a likelihood-based method \citep{Gilbert2006} that relies on three criteria to determine membership: the strength of the \ion{Na}{1} $\lambda\lambda$8190 absorption line doublet, the \textit{(V, I)} color-magnitude diagram location, and photometric versus spectroscopic (\ion{Ca}{2} $\lambda\lambda$8500) metallicity estimates. Following \citet{Gilbert2007}, we excluded radial velocity as a criterion to result in a more complete sample. In total, we identified 37 M31 stellar halo members (20 $\lesssim$ $I_0$ $\lesssim$ 22.5) in this field out of 106 targets.

We required that our abundance measurement technique determined the abundances reliably (\S~\ref{sec:gc_metallicity}): $\delta$([Fe/H]) $<$ 0.5 and $\delta$([$\alpha$/Fe]) $<$ 0.5.  We also required that the 5$\sigma$ $\chi^2$ contours in \deleted{each of the four fitted parameters}\added{$T_\textrm{eff}$, [Fe/H], and [$\alpha$/Fe]} (\S~\ref{sec:spec_abund}) identify the minimum. Both of these criteria effectively mimic a S/N cut (S/N $\gtrsim$ \deleted{8}\added{10} \AA$^{-1}$). Lastly, we manually screened member stars for molecular TiO bands between 7055$-$7245 \AA\ \citep{Cenarro2001,Gilbert2006}, where affected stars exhibit a distinctive pattern. Stars with strong TiO absorption tend to be more metal-rich ([Fe/H] $\gtrsim$ $-$1.5), have red colors (($V$ $-$ $I$)$_{0}$ $>$ 2.0), and can also show unusual $\chi^2$ contours in [$\alpha$/Fe]. We omitted \deleted{3}\added{7} M31 member stars \added{that passed the aforementioned cuts}, which meet the ($V$ $-$ $I$)$_{0}$ color criterion and show spectral evidence of strong TiO absorption. In total, this reduces the sample size to \deleted{14}\added{11} stars (S/N $\sim$ \deleted{8$-$22}\added{10$-$30} \AA$^{-1}$), for which we present a summary of stellar parameters and chemical abundances in Table~\ref{tab:m31_data}.

\added{In Figure~\ref{fig:cmd}, we show the ($I$, $V-I$) color-magnitude diagram for all 37 M31 RGB stars in f130\_2, highlighting the stars contained in our final sample. No stars in our final sample have $(V-I)_0$ $>$ 2.0 due to the aforementioned exclusion of stars with TiO absorption. Excluding this known color bias, the final sample of RGB stars is well-sampled from the full color-magnitude distribution of M31 member stars for this field.} 

\subsection{Kinematics}

\begin{figure}
    \centering
    \includegraphics[width=\columnwidth]{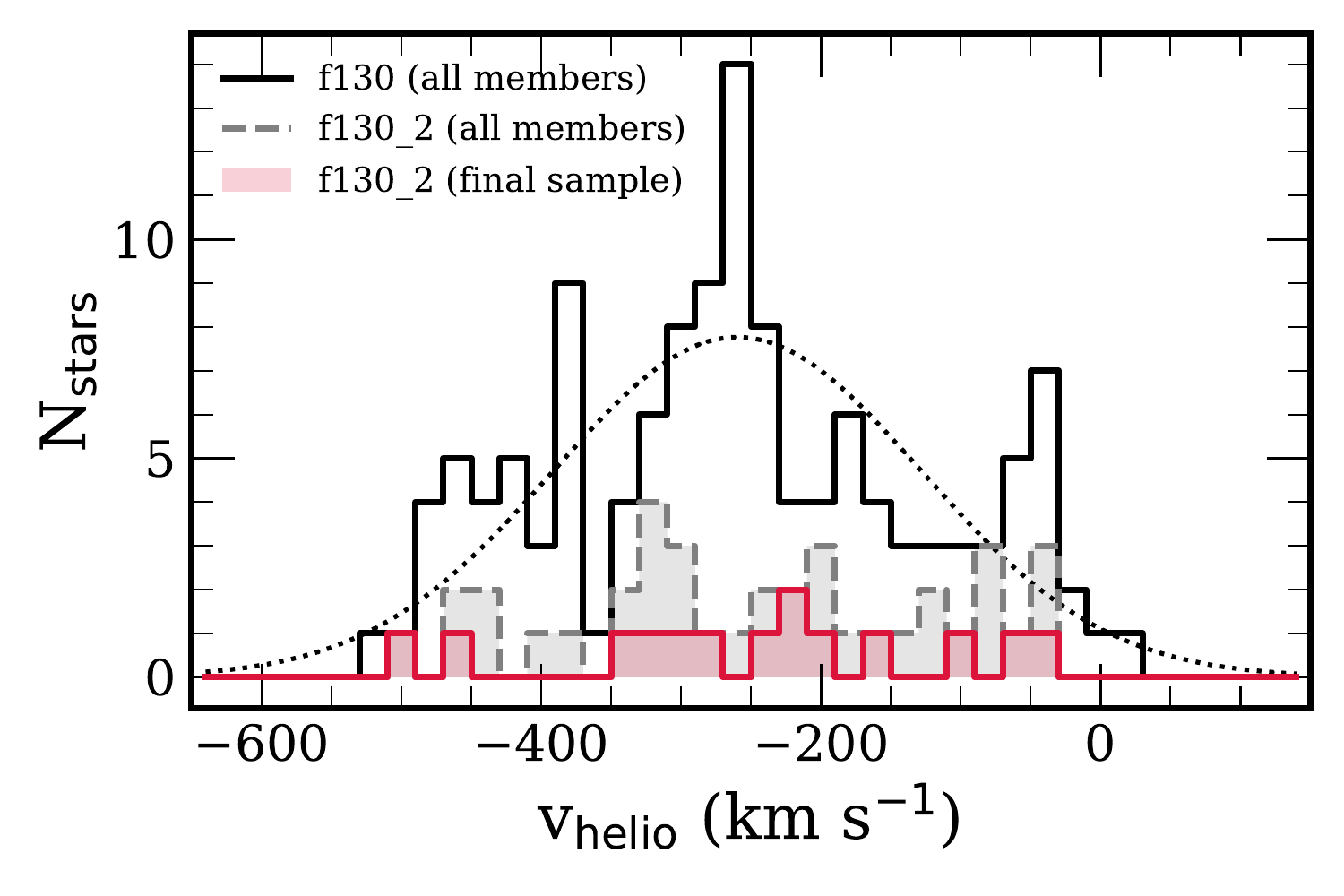}
    \caption{\added{Heliocentric velocity histogram for the final 11 star sample drawn from f130\_2 (\textit{red}) compared to the distributions for all 37 M31 RGB stars in f130\_2 (\textit{grey}) with successful radial velocity measurements. We also show the velocity histogram for 128 M31 RGB stars (\textit{black}) from the broader sample of nearby fields, including f130\_2, known as f130 \citep{Gilbert2007}. The dotted line is the best-fit Gaussian ($\bar{v} = -260$ km s$^{-1}$, $\sigma_v = 132$ km s$^{-1}$; \citealt{Gilbert2007}) to f130, which corresponds to a kinematically hot spheroid component with no detected substructure (i.e., the smooth stellar halo of M31). We find that our final sample is consistent with the kinematics of the hot spheroid.}}
    \label{fig:vhelio_f130_2}
\end{figure}

\begin{figure}
    \centering
    \includegraphics[width=\columnwidth]{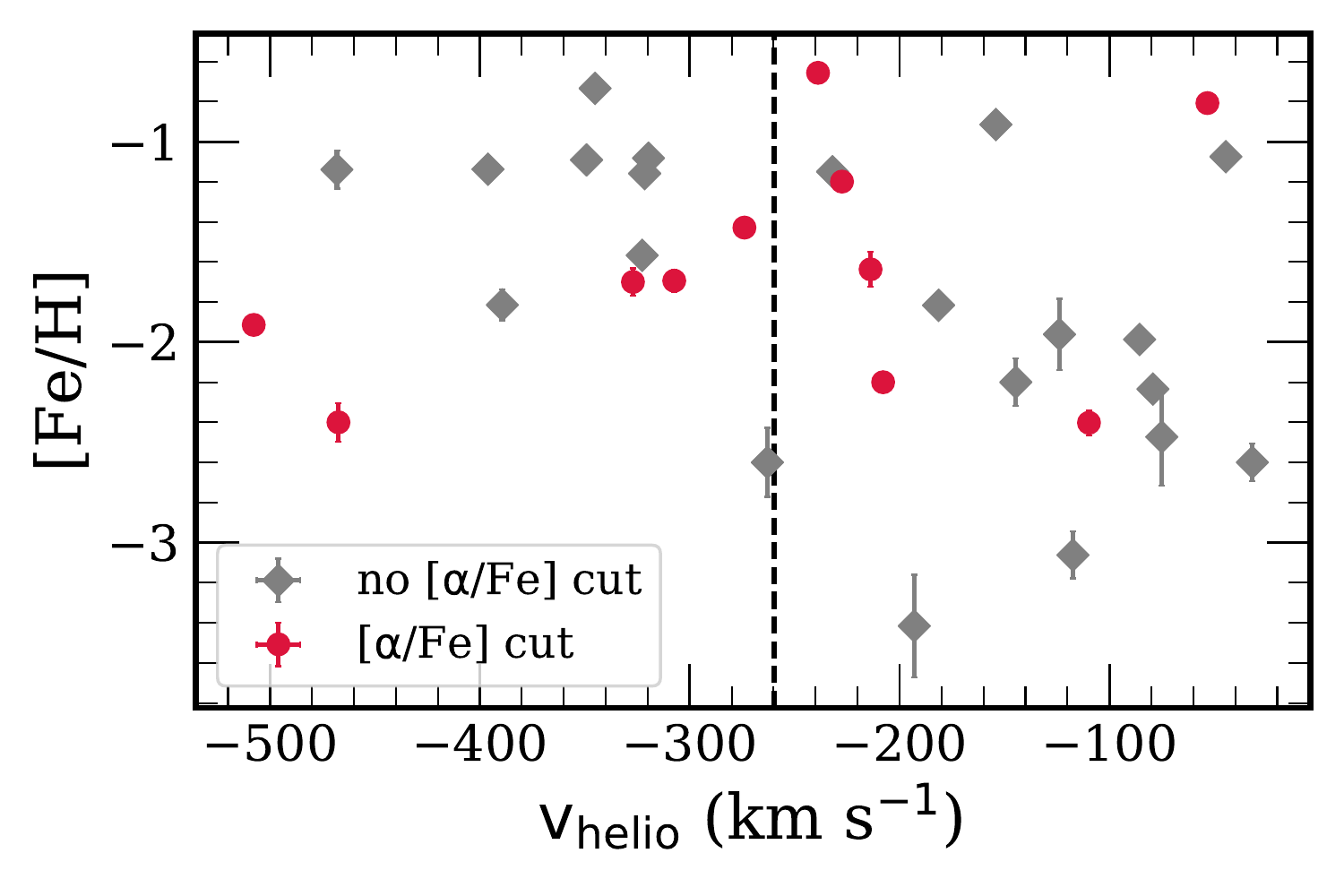}
    \caption{\added{Spectroscopic metallicity ([Fe/H]) as a function of heliocentric velocity for M31 RGB stars in f130\_2. The typical uncertainty in velocity ($\sim$3.5 km s$^{-1}$) is smaller than the size of the data points. In addition to our final sample (\textit{red circles}), which contains only reliable [$\alpha$/Fe] measurements (\S~\ref{sec:m31_sample}), we show a broader sample (\textit{grey diamonds}) that contains well-constrained [Fe/H] measurements, with no cuts on [$\alpha$/Fe]. No apparent correlation exists between radial velocity and metallicity in the field, where our final sample is representative of the broader sample.}}
    \label{fig:vhelio_vs_feh_f130_2}
\end{figure}

\added{Given the proximity of our 23 kpc field to the various structures present in the inner halo of M31,  we analyzed the kinematics of our finalized 11 star sample relative to that of the broader field. We adopt the heliocentric velocity measurements of \citet{Gilbert2007}, which are based on $\sim$1 hour observations obtained with the DEIMOS 1200 line mm$^{-1}$ grating. As discussed in detail by \citet{Gilbert2007}, f130\_2 contains no detectable substructure and is consistent with the kinematics of a hot stellar halo. Additionally, f130\_2 is not a significant contributor to the nearby $-$300 km s$^{-1}$ kinematically cold component known as the Southeast shelf \citep{Gilbert2007,Fardal2007}. Neither is the field spatially coincident with this feature, given that it is located at a larger minor-axis distance (23 kpc) than its outermost extent in projected radius (18 kpc). 

In order to confirm that our final sample is not biased in radial velocity, we present its velocity distribution compared to that of all 37 M31 RGB stars with successful radial velocity measurements in the field in Figure~\ref{fig:vhelio_f130_2}. We also show the velocity distribution for 128 M31 RGB stars from the more encompassing field f130  \citep{Gilbert2007}, composed of 3 distinct slitmasks, including f130\_2. Based on a two-sided Kolmogorov-Smirov test, our final sample is consistent at the 99\% level with the kinematics of a hot spheroid representing the velocity distribution of f130 ( $\bar{v} = -260$ km s$^{-1}$, $\sigma_v$ = 132 km s$^{-1}$; \citealt{Gilbert2007}). We identify this kinematic component with the virialized stellar halo of M31.

We also investigate whether f130\_2 contains any chemically distinct stellar populations. Figure~\ref{fig:vhelio_vs_feh_f130_2} illustrates the relationship between [Fe/H] and radial velocity for our final sample. For a more complete representation of these two quantities, we identified stars in f130\_2 that possessed well-constrained [Fe/H] measurements (\S~\ref{sec:m31_sample}), without enforcing any criteria on the quality of the [$\alpha$/Fe] measurements. We do not find compelling evidence for correlations between [Fe/H] and radial velocity for f130\_2, such that we conclude that there are no kinematically or chemically distinguishable stellar populations within this field. The presence of a tidal feature is not necessary to explain the metallicity or velocity distribution in this field, which is fully consistent with a virialized, phase-mixed stellar population. This conclusion is supported by inspection of the color-magnitude diagram (Figure~\ref{fig:cmd}) and the velocity distribution for the broader field (Figure~\ref{fig:vhelio_f130_2}).

We acknowledge the possibility that kicked-up M31 disk stars, which are kinematically indistinguishable from halo stars \citep{Dorman2013}, could contribute to f130\_2. However, given the distance of f130\_2 along the minor axis, it is unlikely that this fraction exceeds $\sim$1\% \citep{Dorman2013}. Based on our above kinematic analysis, we can rule out any significant contribution to f130\_2 ($\sim$100 kpc in the disk plane) by the extended disk of M31 \citep{Ibata2005} (outermost extent $\sim40-70$ kpc in the disk plane). Thus, we conclude that our final sample of 11 M31 RGB stars accurately represents the properties of the smooth stellar halo of M31 in this region.}

\subsection{Results and Interpretation}
\label{sec:m31_results}

\begin{table*}
\centering
\begin{threeparttable}
\caption{Parameters of 11 M31 RGB Stars}
\begin{tabular}{lcccccc}
\hline
\hline
Object & T$_\textrm{eff}$\tnote{a} (K) & $\log$ $g$ (dex) & [Fe/H] (dex) & [$\alpha$/Fe] (dex) & $\Delta\lambda$\tnote{a} (\AA) & S/N (\AA$^{-1}$)\\
\hline
1282178 & 4339 $\pm$ 7 & 0.39 & -2.4 $\pm$ 0.17 & 0.4 $\pm$ 0.32 & 2.79 $\pm$ 0.04 & 26\\
1292468 & 3796 $\pm$ 4 & 0.67 & -0.66 $\pm$ 0.14 & 0.52 $\pm$ 0.37 & 2.75 $\pm$ 0.03 & 12\\
1292496 & 4368 $\pm$ 5 & 0.7 & -0.81 $\pm$ 0.14 & -0.0 $\pm$ 0.27 & 2.86 $\pm$ 0.02 & 24\\
1292507 & 3899 $\pm$ 5 & 0.46 & -1.69 $\pm$ 0.15 & 0.61 $\pm$ 0.35 & 2.72 $\pm$ 0.05 & 19\\
1302682 & 4075 $\pm$ 5 & 0.85 & -1.43 $\pm$ 0.14 & 0.83 $\pm$ 0.18 & 2.79 $\pm$ 0.02 & 19\\
1302710 & 4264 $\pm$ 9 & 1.07 & -1.64 $\pm$ 0.16 & 0.4 $\pm$ 0.4 & 2.83 $\pm$ 0.04 & 10\\
1302971 & 3858 $\pm$ 3 & 0.53 & -1.7 $\pm$ 0.15 & 0.75 $\pm$ 0.39 & 2.79 $\pm$ 0.04 & 15\\
1303039 & 4144 $\pm$ 4 & 0.52 & -2.2 $\pm$ 0.15 & -0.07 $\pm$ 0.25 & 2.83 $\pm$ 0.02 & 24\\
1303200 & 4337 $\pm$ 4 & 0.88 & -1.91 $\pm$ 0.15 & 0.07 $\pm$ 0.28 & 2.83 $\pm$ 0.02 & 26\\
1303382 & 4356 $\pm$ 3 & 0.78 & -2.4 $\pm$ 0.15 & 0.7 $\pm$ 0.19 & 2.79 $\pm$ 0.02 & 30\\
1303502 & 3914 $\pm$ 3 & 0.39 & -1.2 $\pm$ 0.14 & 0.53 $\pm$ 0.15 & 2.96 $\pm$ 0.02 & 28\\
\hline
\end{tabular}
\begin{tablenotes}
\item[a] As discussed in \S~\ref{sec:sys}, the errors presented for T$_\textrm{eff}$ (and $\Delta\lambda$) represent only the random component of the total uncertainty.
\end{tablenotes}
\label{tab:m31_data}
\end{threeparttable}
\end{table*}

Our \deleted{14}\added{11} measurements increase the previous sample size for [$\alpha$/Fe] measurements in the stellar halo of M31 from 4 stars \citep{Vargas2014b}. For our field, we find inverse-variance weighted values of $\langle$[Fe/H]$\rangle$ = $-$\deleted{1.53}\added{1.59} dex (for a comparison to previous work, see Appendix~\ref{sec:mean_feh}), $\sigma$([Fe/H]) = \deleted{0.52}\added{0.56} dex, $\langle$[$\alpha$/Fe]$\rangle$ = 0.49 dex, and $\sigma$([$\alpha$/Fe]) =  \deleted{0.31}\added{0.29} dex for our uniform, $\alpha$-enhanced halo field at 23 kpc.
%Figure~\ref{fig:f130_2_mgt} illustrates our ability to measure [Fe/H] and [$\alpha$/Fe] using spectral synthesis for one of the highest S/N stars in our data set (S/N = 22.5 \AA$^{-1}$), where we focus on the uncertainty in the [$\alpha$/Fe] determination. 

\begin{figure}
\centering
\includegraphics[width=\columnwidth]{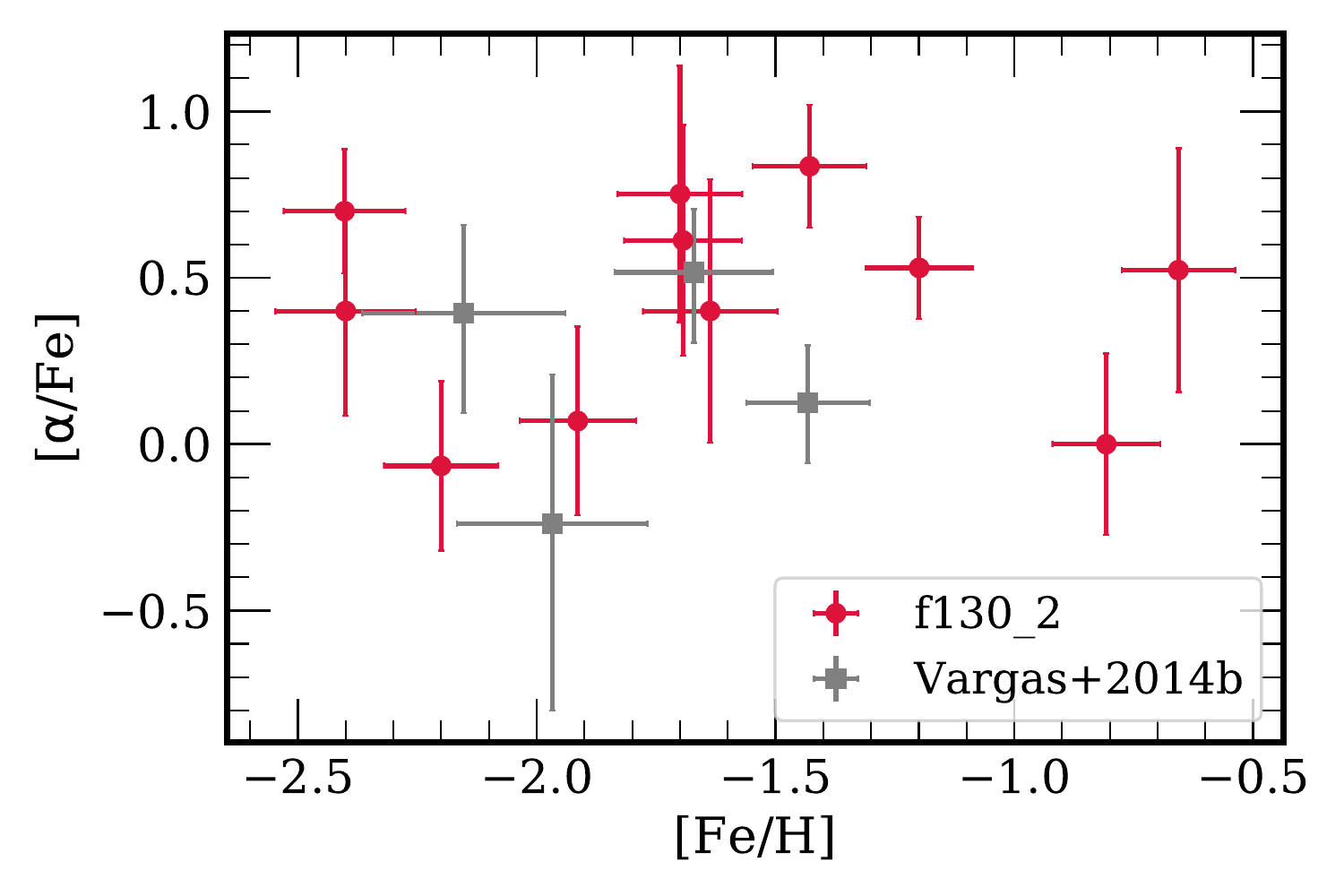}
\caption{[$\alpha$/Fe] vs. [Fe/H], measured from LRS, for M31 RGB stars (\textit{red \added{circles}}) \deleted{from a} \added{in the} 23 kpc field \deleted{with no identified substructure}. We show a subset of our entire sample, with $\delta$([Fe/H]) < 0.5 dex and $\delta$([$\alpha$/Fe]) < 0.5 dex, considering only stars with reliable abundance measurements. In total, we present [$\alpha$/Fe] measurements for \deleted{14}\added{11} M31 halo stars, increasing the previous sample size of 4 stars \citep{Vargas2014b}. We plot the latter sample of metal-poor halo stars (\textit{grey \added{squares}}) (S/N $\gtrsim$ 15 \AA$^{-1}$) over our data set for comparison (S/N $\sim$ \deleted{8$-$22}\added{10$-$30} \AA$^{-1}$). \label{fig:alphafe_vs_feh_m31}} 
\end{figure}

In addition to our \deleted{14}\added{11} measurements of [$\alpha$/Fe] and [Fe/H], Figure~\ref{fig:alphafe_vs_feh_m31} includes the 4 outer halo stars from \citet{Vargas2014b} for comparison. \citet{Vargas2014b} utilized \citeauthor{Gilbert2012}'s (\citeyear{Gilbert2012}) sample of M31 halo stars to identify stars within existing M31 dSph fields \citep{Vargas2014a} for deeper spectroscopic follow-up. They narrowed their sample by enforcing the criteria that the stars were high-likelihood M31 members with S/N sufficient to measure abundances from MRS (S/N $\gtrsim$ 15 \AA$^{-1}$). Their finalized sample originates from the metal-poor outer halo of M31 between $\sim$70 $-$ 140 kpc. We re-compute the inverse-variance weighted average elemental abundances from their data, finding $\langle$[Fe/H]$\rangle$ = $-$1.70 dex, $\sigma$([Fe/H]) = 0.27 dex, $\langle$[$\alpha$/Fe]$\rangle$ = 0.28 dex, and $\sigma$([$\alpha$/Fe]) = 0.22 dex. In contrast to our work, \citet{Vargas2014b} applied an empirical correction factor to convert between the measured, atmospheric value of [$\alpha$/Fe] and the average [$\alpha$/Fe] calculated from individual $\alpha$-element abundances.
%Alternatively, after approximately removing a correction factor to convert between the atmospheric value of [$\alpha$/Fe] and the average [$\alpha$/Fe] calculated from individual $\alpha$-element abundances \citep{Vargas2014a}, we calculate $\langle$[Fe/H]$\rangle$ = $-$1.70 $\pm$ 0.27 and $\langle$[$\alpha$/Fe]$\rangle$ = 0.23 $\pm$ 0.22 dex for the \citet{Vargas2014b} sample. 

%For the sake of comparison, we include an additional M31 field, f123\_1a (Table~\ref{tab:m31_obs}). We observed f123\_1a for a similar exposure time ($\sim$ 3 hr) as f130\_2 using the 1200 line mm$^{-1}$ grating at a central wavelength of 7800 \AA, OG550 order blocking filter, and 0.8'' slit widths. The field f123\_1a is located 18 kpc in projected radius along the minor axis of M31, straddling the edge of the substructure known as the Southeast shelf \citep{Gilbert2007}, a merger remnant of a dwarf galaxy approaching its fourth pericentric passage \citep{Fardal2007}. We identify members in this field in same manner as f130\_2. 

As expected for a smooth halo field, we do not find evidence for a trend of [$\alpha$/Fe] as a function of [Fe/H], in contrast to the expected abundance pattern (decreasing [$\alpha$/Fe] with [Fe/H]) for fields dominated by a single, recent accretion event (such as the Giant Southern Stream; \citealt{Ibata2001}) or dwarf galaxies. Additionally, the fact that our [$\alpha$/Fe] measurements at 23 kpc are consistent with those at $\sim$70$-$140 kpc (Figure~\ref{fig:alphafe_vs_feh_m31}) over the same metallicity range ($-$2.5 dex $\lesssim$ [Fe/H] $\lesssim$ $-$1.5 dex) suggests the lack of a significant radial trend with [$\alpha$/Fe] in M31 stellar halo fields absent of substructure. We also find that our 23 kpc field is on average 0.2 dex more metal rich than the outer halo \citet{Vargas2014b} measurements (see Appendix~\ref{sec:mean_feh} for a discussion of potential selection effects). In combination with the approximately constant value of [$\alpha$/Fe] with both [Fe/H] and radius, this may indicate that we are probing the same extended halo component, which is metal-poor, $\alpha$-enhanced, and underlies substructure at all radii \citep{Chapman2006,Gilbert2012,Ibata2014}. 

Given the low luminosity of the smooth halo component (L $\sim$ 1.9$\times$10$^8$ L$_\odot$ for [Fe/H]$_\textrm{phot}$ $<$ $-$1.1 dex), \citet{Ibata2014} inferred that it would consist of many low luminosity structures accreted at early times. In terms of star formation history (SFH), high $\alpha$-element abundances indicate that the stellar population in f130\_2 is characterized by rapid star formation and is dominated by the yields of Type II supernovae. Recognizing that the outer regions ($\gtrsim$ 20 kpc) of the stellar halo are most likely formed via accretion \citep{Johnston2008,Cooper2010,Tissera2012}, we infer that the disrupted dwarf galaxies that were the progenitors of this field likely had short SFHs. Their SFHs could have been truncated by accretion onto M31. 

Interestingly, the slightly lower average $\alpha$-element abundance (0.28 dex) of \citet{Vargas2014b} could suggest that the outer halo is composed of progenitors with more extended chemical evolution as compared to the inner halo. If true, this would be in accordance with expectations from hierarchical buildup of the stellar halo \citep{Johnston2008,Font2008}. However, we cannot draw a robust conclusion on this matter given that the average $\alpha$-element abundances, similar to the case of [Fe/H], between \citeauthor{Vargas2014b}'s \citeyear{Vargas2014b} sample and our sample are consistent at the 1$\sigma$ level, which is compounded by limited sample sizes.

Our inferred SFH for f130\_2 qualitatively agrees with the trend derived from deep photometry in a nearby \textit{HST}/ACS field located at 21 kpc along the minor axis. The mask centers of the fields are separated by 6.33 arcmin on the sky, or 1.44 kpc, assuming a distance to both fields of 783 kpc \citep{StanekGarnavich1998}. Using the \citet{Brown2006} method of comparing theoretical isochrones to color-magnitude diagrams, \citet{Brown2007} derived a SFH for the ACS field, assuming [$\alpha$/Fe] = 0. They found a wide range of stellar ages and metallicities, providing support for an accretion origin, as opposed to early monolithic collapse. The field exhibits evidence for an extended SFH, with the majority of stellar ages between $\sim$8$-$10 Gyr, with a small but non-negligible ($\lesssim$ 5\%) population of stars with ages $\lesssim$ 8 Gyr. The wide range of metallicity ($-$\deleted{2.3}\added{2.5} $<$ [Fe/H] $<$ $-$\deleted{0.7}\added{0.5} dex) that we find in this work is consistent with a multiple progenitor hypothesis. 
%However, it is also consistent with a single massive progenitor that quickly self-enriched with a high star formation rate at early times. 
If the nearby ACS field is representative of f130\_2, this implies a composition for f130\_2 of intermediate-age system(s) that had elevated star formation rates, quenched at latest \ $\lesssim$ 8 Gyr ago.

Comparing our average $\alpha$-element abundance to that of other systems, we find that, in general, they are similarly $\alpha$-enhanced.\@ $\langle$[$\alpha$/Fe]$\rangle$ for the 23 kpc M31 halo field agrees with that of M31 GCs (0.37 $\pm$ 0.16 dex) within 20 kpc of the galactic center \citep{Colucci2009}. Additionally, the metal-poor MW halo possesses elevated $\alpha$-element abundance ratios of approximately +0.4 dex \citep{Venn2004,Cayrel2004,Ishigaki2012,Bensby2014}, which is comparable to our result.

Drawing comparisons to M31 dwarf galaxies is less straightforward, given that their average $\alpha$-element abundance varies from approximately solar to highly $\alpha$-enhanced ($\sim$0.5 dex) \citep{Vargas2014a}. This may indicate a range of star formation timescales for these systems, where some are dominated by old stellar populations ($\gtrsim$ 10 Gyr ago) and others possess intermediate-age ($\sim$7-10 Gyr ago) stars, although the systematic uncertainties on their SFHs at early times are large \citep{Weisz2014}. \citet{Vargas2014a} also found M31 dwarf galaxies to vary in terms of their internal [$\alpha$/Fe] vs.\ [Fe/H] abundance patterns, ranging from constant (e.g., And VII; \citealt{Tollerud2012}) to decreasing [$\alpha$/Fe] with respect to [Fe/H] (And V; \citealt{Tollerud2012}). The latter case is in accordance with abundance trends found in MW dwarf spheroidal galaxies \citep{Shetrone2001,Shetrone2003,Tolstoy2003VLT/UVESEvolution,Venn2004,Kirby2009,Kirby2011b} and systems with more extended SFHs. 

In terms of $\alpha$-enhancement and SFH, our field resembles old M31 dSphs, although it is possible that f130\_2 contains intermediate age stars \citep{Brown2007}. \citet{Vargas2014a} inferred that a present-day stellar halo constructed from M31 dwarf galaxies would be metal-rich, where $\langle$[Fe/H]$\rangle$ $\sim$ $-$0.7 dex ($-$1.4 dex) for their full sample (old dwarf galaxies only), with a distinct $\alpha$-element abundance pattern as compared to the MW halo. Given the similarly flat [$\alpha$/Fe]-[Fe/H] trend between f130\_2 and And VII, and the similar $\langle$[Fe/H]$\rangle$ and [Fe/H] range between f130\_2 and old M31 dSphs, it is possible that the progenitors of f130\_2 were composed of systems similar to And VII. In order to meaningfully test if systems similar to present day M31 dwarf galaxies could have contributed to the smooth halo component, or whether the $\alpha$-element abundance pattern of the smooth halo \added{of M31} differs from that of the MW, we would require larger sample sizes across more halo fields.

%Discussion of the errorbars. What is the precision at a given S/N that we expect for our intended M31 data set?

\section{Summary}
\label{sec:summary}

In an effort to increase the amount of available high-quality data in M31, we have developed a method of measuring [Fe/H] and [$\alpha$/Fe] from low-resolution spectroscopy of individual RGB stars.  We applied our technique to a field in M31's smooth stellar halo component. 

The primary advantages of utilizing low-resolution spectroscopy are (1) the substantial increase in wavelength coverage (from $\sim$ 2800 \AA\ with MRS to $\sim$ 4600 \AA\ with LRS) available to constrain the abundances and (2) the accompanying increase in S/N per pixel for the same exposure time and observing conditions. To make spectral synthesis of DEIMOS LRS a reality, we generated a new grid of synthetic spectra spanning 4100 $-$ 6300 \AA \ based on a line list we constructed for bluer optical wavelengths. We find the following results:

\begin{enumerate}
    \item Testing our technique on Galactic GCs, we do not find evidence for any systematic covariance between fitted parameters, such as $T_{\rm eff}$ and [Fe/H]. In light of the the fundamental inhomogeniety of the various HRS samples compared to our LRS data set, our measurements broadly agree with HRS abundances.
    \item Based on the intrinsic dispersion in [Fe/H] and [$\alpha$/Fe] of Galactic GCs with no known abundance variations in Fe, Mg, Ca, or Si, we estimate error floors of $\delta$([Fe/H])$_\textrm{sys}$ = \deleted{0.105}\added{0.111} dex and $\delta$([$\alpha$/Fe])$_\textrm{sys}$ = \deleted{0.039}\added{0.094} dex.
    \item We present measurements for \deleted{14}\added{11} RGB stars of [Fe/H] \textit{and} [$\alpha$/Fe] in the stellar halo of M31, increasing the previous sample size of 4 stars. The field has no identified substructure and is located at 23 kpc in galactocentric projected radius. We find that $\langle$[Fe/H]$\rangle$ = $-$\deleted{1.53}\added{1.59} $\pm$ \deleted{0.52}\added{0.56} dex and $\langle$[$\alpha$/Fe]$\rangle$ = 0.49 $\pm$ \deleted{0.31}\added{0.29} dex for this field.
    \item $\langle$[$\alpha$/Fe]$\rangle$ agrees with the value of the MW halo plateau ($\sim$0.4 dex), M31 GCs, and some $\alpha$-enhanced M31 dwarf galaxies. Our measurements exhibit overlap with previously published [$\alpha$/Fe] measurements for M31 halo RGB stars at larger projected radii (70$-$140 kpc), showing no evidence for a significant radial trend in [$\alpha$/Fe] in our limited sample.
    \item Given its high $\alpha$-enhancement \added{and low metallicity}, we surmise that the smooth halo field is likely composed of disrupted dwarf galaxies with elevated star formation rates and truncated SFHs, accreted early in the formation history of M31.
    %\item Precision at a given S/N. 
\end{enumerate}

In future work, we will measure [Fe/H] and [$\alpha$/Fe] from $\sim$6 hour observations of individual RGB stars in additional M31 halo and tidal stream fields with deep \textit{HST} photometry \citep{Brown2006}, with the goal of deriving chemically-based star formation histories. 

\acknowledgments
The authors thank \added{the anonymous reviewer for a careful reading of this manuscript and providing thoughtful feedback that improved this paper. We also} thank Alis Deason for assistance in line list vetting, Gina Duggan for useful discussions on generating grids of synthetic spectra, Raja Guha Thakurta 
%and Alex Ji 
for help with observations and insightful conversations, and Luis Vargas and Marla Geha for sharing their data for M31 outer halo RGB stars. IE acknowledges support from a Ford Foundation Predoctoral Fellowship and the NSF Graduate Research Fellowship under Grant No.\ DGE-1745301, as well as the NSF under Grant No.\ AST-1614081, along with ENK. KMG and JW acknowledge support from NSF grant AST-1614569. ECC was supported by a NSF Graduate Research Fellowship as well as NSF Grant No.\ AST-1616540. The analysis pipeline used to reduce the DEIMOS data was developed at UC Berkeley with support from NSF grant AST-0071048.

\bibliographystyle{apj}
\bibliography{main.bib}

\appendix
\section{Mean Metallicity: Comparison to Previous Work}
\label{sec:mean_feh}

In this work, we focus on the determination of [$\alpha$/Fe] in M31 stellar halo RGB stars. Given the limited sample size of previously existing equivalent measurements, we can only directly compare our [$\alpha$/Fe] measurements to the \citet{Vargas2014b} sample. However, an extensive body of literature exists on [Fe/H] estimates in the stellar halo of M31, which we discuss in detail here in the context of our measurements.

As presented in \S~\ref{sec:m31_results}, we find $\langle$[Fe/H]$\rangle$ = $-$\deleted{1.53}\added{1.59} dex and $\sigma$([Fe/H]) = \deleted{0.52}\added{0.56} dex for f130\_2. In contrast, \citet{Brown2007} estimated $\langle$[Fe/H]$\rangle_{\rm{phot}}$ = -0.87 dex for the nearby ACS field from color-magnitude diagram based SFHs, where their value is more metal-rich than our mean metallicity by \deleted{0.72}\added{0.71} dex. In terms of both star counts and metallicity, \citet{Brown2007} characterized this field as straddling a transition region between the metal-rich inner halo and the metal-poor outer halo. Although the extended halo ($\gtrsim$ 60 kpc) is known to be metal-poor based on both photometric and Ca triplet metallicity indicators \citep{Guhathakurta2006,Kalirai2006b,Chapman2006,Koch2008,Gilbert2014,Ibata2014}, a majority of photometric studies find that the inner halo (20$-$30 kpc) is as metal-rich as $-$0.7 dex for fields unpolluted by Giant Southern Stream debris \citep{Guhathakurta2006,Gilbert2014}. Based on an imaging survey, \citet{Ibata2014} found [Fe/H]$_\textrm{phot}$ = $-$0.7 dex at 30 kpc for [$\alpha$/Fe] = 0, where the mean metallicity does not decline to $-$1.5 dex until 150 kpc. Assuming [$\alpha$/Fe] = 0.3 dex, \citet{Kalirai2006b} found $\langle$[Fe/H]$\rangle_\textrm{phot}$ = $-$1.48 dex and $\sigma$([Fe/H]$_\textrm{phot}$) = 0.11 dex for the extended metal-poor halo ($\gtrsim$ 60 kpc). They based their measurements on photometry from fields with $\sim$1 hour DEIMOS spectroscopy, but they did not include f130\_2 in their analysis of inner halo fields, for which they found $\langle$[Fe/H]$\rangle_\textrm{phot}$ = $-$0.94 dex and $\sigma$([Fe/H]$_\textrm{phot}$) = 0.60 dex around 30 kpc. Similarly, based on 397 stars between 20$-$40 kpc, \citet{Gilbert2014} found $\langle$[Fe/H]$\rangle_\textrm{phot}$ = $-$0.70 dex and $\sigma$([Fe/H]$_\textrm{phot}$) = 0.53 dex for [$\alpha$/Fe] = 0 in this region (including more metal-rich Giant Southern Stream debris).

Clearly, our value of $\langle$[Fe/H]$\rangle$ = $-$\deleted{1.53}\added{1.59} dex for f130\_2 is discrepant with photometric studies of M31's inner halo. This could be a consequence of selection effects against metal-rich stars, given that we discarded stars with strong TiO absorption (\S~\ref{sec:m31_sample}). However, we also consider alternative explanations. There are indications that (1) a smooth, metal-poor halo component with no detected substructure is found at all radii, and (2) the photometric metallicities likely overestimate the degree to which the inner halo is metal-rich. Using Ca triplet equivalent width measurements from stacked DEIMOS spectra, \citet{Chapman2006} analyzed major axis fields (and one minor axis field) in M31's stellar halo, finding evidence for a metal-poor stellar halo component ([Fe/H]$_\textrm{CaT}$ = $-$1.4 dex) detectable at all radii between 10$-$70 kpc with no apparent metallicity gradient. In an analysis of M31's surface brightness profile, \citet{Gilbert2012} confirmed the detection of this distinct halo component. Additionally, \citet{Ibata2014} found that the smooth halo is $\sim$0.2 dex more metal-poor than fields dominated by substructure, where metallicities of $-2.5$ $<$ [Fe/H] $<$ $-1.1$ tend to characterize fields throughout the halo with little to no substructure. In contrast to \citet{Kalirai2006b}, \citet{Koch2008} analyzed the same DEIMOS fields (including f130\_2) using Ca triplet metallicities, finding values systematically more metal-poor in mean metallicity by $\sim$0.75 dex. The large discrepancy likely results from differences in sample selection and metallicity measurement methodology (photometric vs. Ca triplet based).

%At a projected radius of 20$-$40 kpc they found [Fe/H]$_\textrm{CaT}$ $\sim$ $-$1.4 dex, which is in better agreement with the value found in this work. They also found that fields along the minor axis were more metal-poor at a given radius than off-axis fields, which were more likely to be polluted by the Giant Southern Stream.

%In response to the latter study, \citet{Brown2007} noted that if the \citet{Koch2008} metallicity trend is more representative of the halo, the \textit{HST}/ACS fields could be in regions of systematically higher [Fe/H]. Although this may be true for ACS fields that probe substructure, a more likely explanation for the apparent discrepancy is the large uncertainties of photometric metallicties \citep{Lianou2011}, which tend to over-estimate the metallicity of metal-poor stars. 

Whether the methodology employed is photometric, Ca triplet based, or utilizes spectral synthesis can result in substantial differences in metallicity estimates for the same sample (e.g., \citealt{Lianou2011}).
%In particular, photometric estimates tend to have large uncertainties and overestimate the metallicity of metal-poor stars, where discrepancies with spectroscopic metallicities increase substantially (as high as $\sim$0.5 dex) for systems with intermediate-age stellar populations (e.g., \citealt{Lianou2011}). The assumption of a uniformly ancient stellar population does not hold in the stellar halo of M31 \citep{Brown2006,Brown2007,Brown2008}, thus single-age isochrones likely cannot provide reliable metallicity estimates.
Most relevantly, photometric studies often assume [$\alpha$/Fe] = 0, which can inflate metallicity estimates significantly compared to assuming an $\alpha$-enhanced field. 
%\citet{Gilbert2014} found that assuming [$\alpha$/Fe] = 0.3 dex, as opposed to [$\alpha$/Fe] = 0, resulted in a $\sim$0.17 dex decrease in [Fe/H]$_\textrm{phot}$. 
Using \citet{VandenBerg2006} isochrones, assuming 10 Gyr old stellar populations \citep{Brown2007}, a distance modulus of $(m - M)_0$ = 24.63 $\pm$ 0.20 \citep{Clementini2011}, and [$\alpha$/Fe] = 0 dex, we found $\langle$[Fe/H]$\rangle_\textrm{phot}$ = $-$1.40 dex for our sample of \deleted{14}\added{11} M31 RGB stars. If we instead assume [$\alpha$/Fe] = 0.3, we obtain $\langle$[Fe/H]$\rangle_\textrm{phot}$ = $-$1.60 dex, corresponding to a decrease in the mean photometric metallicity of 0.19 dex. We find nearly identical results by repeating the calculation with a different set of isochrones \citep{Demarque2004}. 

The assumptions intrinsic to photometric metallicities, combined with the large amount of tidal debris present in the inner halo of M31 that is included in many previously published measurements in this radial range, are sufficient to explain the large difference between our value of $\langle$[Fe/H]$\rangle$ for f130\_2 and previous analyses in the inner halo of M31. A primary strength of our study is that we can determine both [$\alpha$/Fe] and [Fe/H] from spectroscopy, without prior assumptions on either parameter. We acknowledge that we may be preferentially sampling brighter, more metal-poor stars in this field, given that we are S/N-limited and select against stars with strong TiO absorption. However, given that we can measure both [Fe/H] and [$\alpha$/Fe] reliably from some of the highest quality spectra in M31's halo yet obtained, we conclude that $\langle$[Fe/H]$\rangle$ = $-$\deleted{1.53}\added{1.59} dex is likely an accurate representation of our final sample's mean metallicity. Thus, it is possible that our sample in f130\_2 represents the metal-poor halo that underlies substructure \citep{Chapman2006,Gilbert2014,Ibata2014} in the inner halo of M31.

\end{document}